\documentclass[journal]{IEEEtran}
\usepackage{amsmath,amsfonts}
\usepackage{algorithmic}
\usepackage{array}
\usepackage{textcomp}
\usepackage{stfloats}
\usepackage{url}
\usepackage{verbatim}
\usepackage{graphicx}
\usepackage{cite}
\usepackage{subcaption}
\usepackage{color}
\usepackage{siunitx}
\usepackage{amsmath,amssymb}
\usepackage{threeparttable}
\usepackage{adjustbox}

\usepackage{arydshln}
\usepackage{svg}
\newcommand{\orcid}[1]{\href{https://orcid.org/#1}{\includesvg[width=10pt]{orcid}}}

\usepackage[bookmarks=false]{hyperref} 

\begin{document}
\title{Compact Dual-Polarization Schottky Barrier Diode Receivers for Submillimeter Wave Remote Sensing}

\author{Olivier~Auriacombe,~
    Peter J.~Sobis,~
    Vladimir~Drakinskiy,~\IEEEmembership{Member,~IEEE,}
    Anders~Emrich,~
    and~Jan~Stake,~\IEEEmembership{Fellow,~IEEE}

\thanks{Manuscript received May 18\textsuperscript{th}, 2026, revised July X, 2026. This research was in part funded by the European Space Agency ('Dual-probe Terahertz receiver' 2017-2019 and 'Integrated Dual Polarisation 650/325~GHz Receivers' 2021-2023), and was carried out in the Advanced Digitalization program at the WiTECH Centre CRYTER+, which has been financed by VINNOVA, Chalmers University of Technology, Omnisys Instruments AB (AAC Omnisys), Low Noise Factory AB (LNF), Research Institutes of Sweden (RISE) and Virginia Diodes Inc. (VDI).}
\thanks{Olivier Auriacombe is with Omnisys Instruments AB and the Department of Microtechnology and Nanoscience, Chalmers University of Technology, Sweden. (e-mail: \mbox{oliaur@chalmers.se})}
\thanks{Peter Sobis was with Omnisys Instruments AB and Chalmers University of Technology, SE-412 96 Gothenburg, Sweden. He is now with Low Noise Factory AB, Gothenburg, Sweden.}
\thanks{Anders Emrich was with Omnisys Instruments AB. He is now with EmriX Consulting AB, Askim, Sweden.}
\thanks{Jan Stake and Vladimir Drakinskiy are with the Department of Microtechnology and Nanoscience, Chalmers University of Technology, SE-412 96 Gothenburg, Sweden. (e-mail: \mbox{jan.stake@chalmers.se})}
}

\markboth{IEEE Transactions on Terahertz Science and Technology,~Vol.~x, No.~x, X~2026}%
{How to Use the IEEEtran \LaTeX \ Templates}

\maketitle

\begin{abstract}
Dual-polarization heterodyne receivers operating at 325~GHz, 424~GHz, and 650~GHz at room temperature are presented. Polarimetric measurements are enabled by two orthogonal open-ended \textit{E}-field probes, co-optimized and integrated with two subharmonic GaAs Schottky-barrier diode mixers. The down-converted signals (IF) are amplified using low-noise InP HEMT amplifiers integrated into the receiver module, along with IF matching networks, dc-bias boards, a shared local oscillator (LO) distribution network, and a single smooth-walled, conical, spline-horn antenna. Maximum cross-polarization isolation of 25~dB, 34~dB, and 25~dB was achieved at 315~GHz, 421~GHz, and 650~GHz, respectively. The measured double-sideband (DSB) receiver noise temperatures are 800~K, 800~K, and 1600~K at 315~GHz, 421~GHz, and 630~GHz, respectively. Stability measurements, with an integration Allan time of more than 10 s, were obtained for all receivers. Overall, the integrated dual-polarization receiver topology achieves excellent sensitivity in a highly compact package, offering an efficient and scalable solution for polarimetric applications in submillimeter-wave remote sensing.
\end{abstract}


\section{Introduction}
\IEEEPARstart{P}{olarimetry} in the millimeter and submillimeter wave region \cite{Siegel2002} is a widely used technique in Earth remote sensing to retrieve ice cloud particle properties and surface emissivity (e.g., in forestry, hydrology, and crop monitoring), as well as in radio astronomy to investigate the polarization characteristics of molecular emissions. By simultaneously measuring the horizontal (H) and vertical (V) polarized brightness temperatures of a target scene, all four Stokes parameters \cite{Walker1954} can be retrieved, providing additional information such as sea surface wind speed and direction, or the orientation distribution and crystal shapes of hydrometeors in the atmosphere \cite{Kaur2022}. Direct measurements of the total mass of ice hydrometeors \cite{Eriksson2018} are scarce within the atmospheric hydrological cycle. Using submillimeter-wave sensing could enhance numerical weather prediction by providing better constraints on hydrometeor properties for all-sky data assimilation. Additionally, dual-polarization measurements are used in communication and various remote sensing applications, including terahertz ellipsometry \cite{Kuhne2018}, imaging \cite{Shi_2023},  millimeter-wave road-surface emissivity measurements \cite{Auriacombe2025}, and ground-based tropospheric temperature sounding and Zeeman-effect studies \cite{Krochin2022}. Hence, efficient, miniaturized dual-polarization receivers are crucial for fully exploiting polarimetric observations in the microwave and submillimeter wavelength range. 

Since the early 2000s, several spaceborne submillimeter passive receivers have been launched \cite{Thomas2025}, such as the Odin satellite \cite{Murtagh2002} and NASA’s EOS MLS aboard the Aura satellite \cite{Waters2006}. A precursor instrument for hydrometeor measurements is the ISMAR \cite{Fox2017}, an airborne demonstrator capturing dual-polarization brightness temperatures at 240~GHz, 664~GHz, and 874~GHz \cite{Hammar2018}. In May 2017, the first spaceborne 883~GHz radiometer was launched from the ISS on a CubeSat \cite{Gong2021}. Recent developments include the Arctic Weather Satellite (AWS) \cite{Eriksson2025}, a cross-track microwave sounder launched in 2024 to improve weather forecasting, and NASA TROPICS \cite{Blackwell2025}, a satellite constellation providing high-resolution atmospheric observations from microwave sounders in low-Earth orbit. Further submillimeter-wave sounders are under development to improve observations of ice clouds above 325~GHz. For example, NASA’s compact submillimeter-wave and long-wave infrared polarimeters for cirrus ice properties (SWIRP) instrument comprises two dual-polarization channels operating at 220~GHz and 680~GHz \cite{Cooke2020}, while the Polarized Submillimeter Ice-cloud Radiometer (PolSIR) concept involves two dual-polarised channels at 325~GHz and 680~GHz \cite{Wu2024}. Hence, microwave sounders and imagers are essential for atmospheric sciences. In particular, three key frequency bands have been identified to characterize ice particle sizes and improve understanding of cloud microphysics. The 325~GHz band is beneficial for retrieving the effective radius of ice particles, particularly in cirrus clouds. The 448~GHz band provides sensitivity to the ice water path and high-altitude cloud properties, while the 650~GHz band, with its high sensitivity to smaller ice particles, enhances radiative transfer modelling.  In 2026, ESA’s Ice Cloud Imager (ICI) will be launched as part of the EUMETSAT MetOp-SG-B (Second Generation B) operational satellite series. ICI will measure atmospheric ice mass at 325.15~GHz~($\pm$ 1.5, 3.4, 9.5~GHz), 448~GHz~($\pm$ 1.4, 3.0, 7.2~GHz), and 664~GHz~($\pm$ 4.2~GHz), with the 243~GHz and 664~GHz channels involving both polarizations \cite{Eriksson2020}. Thus, there is a pressing need to develop compact, efficient, and sensitive polarimetric receivers for future small-satellite platforms \cite{Bryerton2016}.

At microwave frequencies, H- and V-polarizations are typically acquired using traditional waveguide-based orthomode transducers (OMTs) \cite{Boifot1990}. However, waveguide losses increase with frequency \cite{Skinner2025}, degrading radiometric performance \cite{Dicke1946}. In 2013, Reck and Chattopadhyay demonstrated an asymmetric waveguide OMT at 600~GHz \cite{Reck2013} with a return loss of 20~dB for both polarizations and a polarization isolation better than 25~dB. Silicon micromachining techniques have been introduced to reduce losses and improve fabrication tolerances with state-of-the-art performance for OMTs in the 220-330-GHz band \cite{GomezTorrent2019}. Alternatively, to overcome the complexity of OMTs at terahertz frequencies, quasi-optical techniques are often employed to co-locate multiple beams \cite{Woodcraft2023}, offering broader bandwidth and lower loss, though at the expense of increased system complexity and size.   A promising alternative uses orthogonal RF probe transitions, first proposed for a rectangular waveguide in 1979 \cite{JOGLEKAR1979}, enabling planar integration \cite{Jackson2001}. Engargiola and Plambeck showed that four planar probes in a circular waveguide, with each pair connected to a balun, can provide cross-polarization isolation better than 35~dB at 1.5~GHz \cite{Engargiola2003}.  Nonetheless, scaling and integrating four probes and two baluns at shorter wavelengths pose significant challenges. However, Liu \textit{et al.} \cite{KuanYuLiu2013} reported over 20-dB cross-polarization isolation across the 306-367~GHz band in a dual-polarization SIS mixer with two planar circular-waveguide probes, which is more than sufficient for many applications. 

In this work, we present integrated dual-polarization Schottky-diode-based heterodyne receivers that employ two orthogonal, asymmetric, open-ended \textit{E}-field probes in a circular waveguide \cite{Sobis2018}, fed by a single spline-profile conical horn antenna. Such compact receivers meet the need for state-of-the-art radiometric performance in a reduced physical envelope reserved for receivers and optical networks, allowing their deployment in small satellites or CubeSats. In addition, a dual-polarization receiver module provides redundancy in case of channel failure without incurring penalties in additional electrical power or mechanical space. Based on initial results presented in 2022 \cite{Auriacombe2022}, three receiver modules are designed and demonstrated at operating frequencies of 325~GHz, 424~GHz, and 650~GHz, respectively. This integrated circuit enables direct, simultaneous measurement of orthogonal polarization components using an on-chip probe and downconversion via heterodyne mixing on the same substrate. The design eliminates the need for quasi-optical beam separation while avoiding the high losses associated with conventional waveguide OMTs \cite{Skinner_1991}. The design and experimental demonstration are described in the following sections.

\section{Dual-polarization receiver}
\subsection{Integrated Receiver Overview}
\begin{figure}[ht!]
    \centering
    \includegraphics[width=1\linewidth]{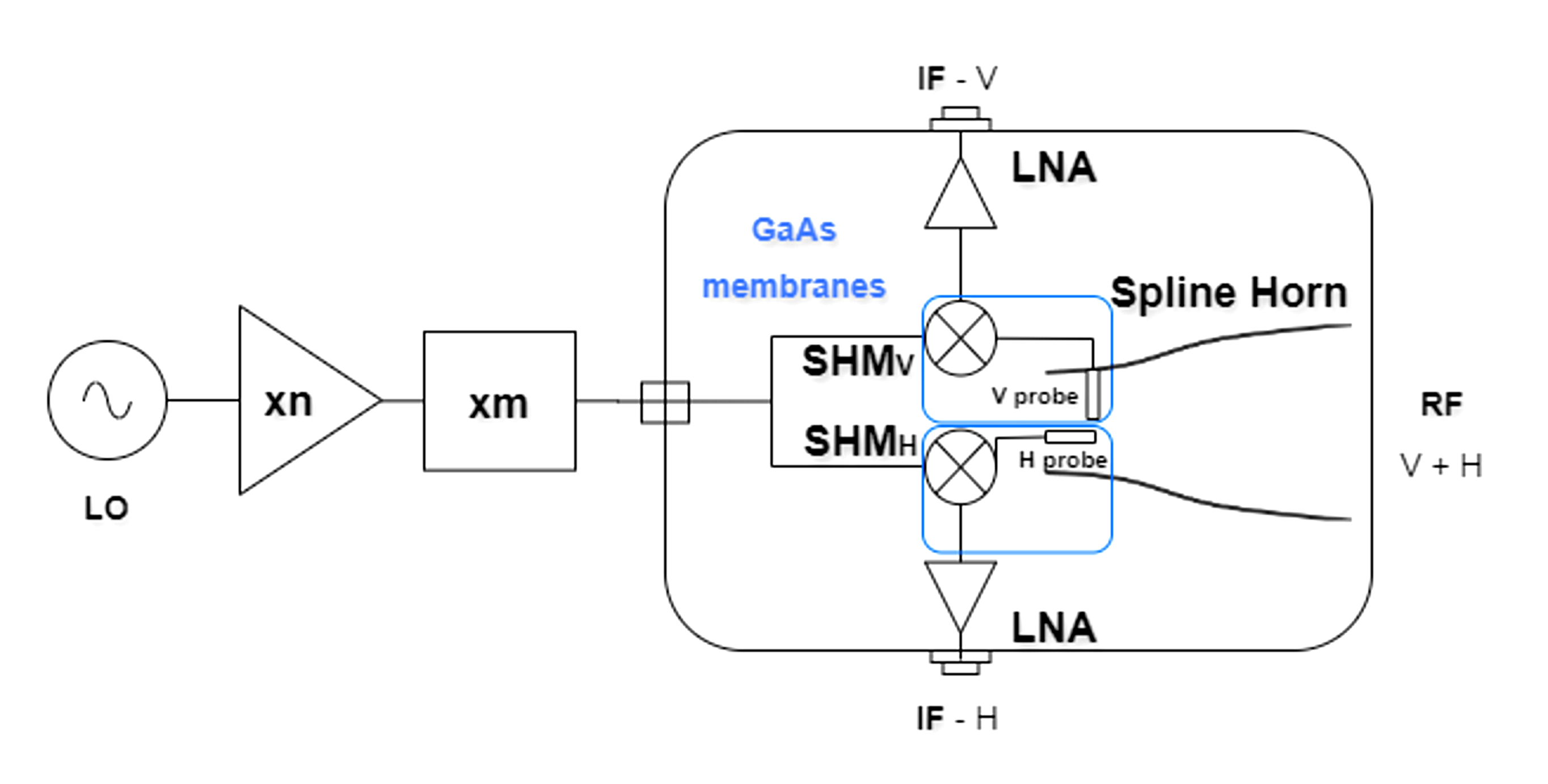}
    \caption{Schematic of the dual-polarization receiver architecture.}
    \label{fig:receiver_schematic}
\end{figure}

The complete polarimetric receiver is integrated into a single mechanical split-block assembly that features a single RF horn antenna, see Fig. \ref{fig:receiver_schematic}. Both receiver channels use planar orthogonal probes to extract the polarization components of the incoming RF signal in a circular waveguide \cite{KuanYuLiu2013}. Each \textit{E}-field probe is integrated with the Schottky barrier diode mixer circuit \cite{Mehdi2017} on a $3~\mu$m-thick gallium arsenide (GaAs) substrate, and the down-converted intermediate frequency (IF) signal is amplified using a low-noise amplifier (LNA). The low RF path losses between the horn aperture and the Schottky barrier diodes, together with the compact system design, help achieve a state-of-the-art receiver noise temperature.

 A reference local oscillator (LO) generates a tone at around 10~GHz for the 325-GHz and 650-GHz receivers. A reference local oscillator (LO) generates a tone at around 10~GHz for the 325-GHz and 650-GHz receivers based on the LMX2595 synthetiser chip with very low phase noise. This permits simplicity in instrument design and eliminates potential interference from many LO sources. The LO tone is generated at 35~GHz for the 424-GHz receiver. Active frequency multipliers (AMC-12-RNHB2) are employed in the 325~GHz and 650~GHz LO chains. These are followed by Schottky-diode-based frequency doublers (Radiometer physics GmbH) that up-convert the LO signals to 162.5~GHz and 325~GHz, respectively, for injection into the 325-GHz and 650-GHz dual-polarization receivers. The LO chain for the 424-GHz receiver utilizes a high-power, heterostructure barrier varactor (HBV), WR-10 tripler \cite{Malko2012} from Wasa Millimetre Wave AB, followed by a Schottky varactor doubler developed at Chalmers University of Technology \cite{Hammar2018}. The frequency multiplication coefficients $n$ and $m$ in Fig.~\ref{fig:receiver_schematic} correspond to $\times8$ ($n$) and $\times2$ ($m$) for the 325-GHz receiver, $\times1$ ($n$) and $\times6$ ($m$) for the 424-GHz receiver, and $\times8$ ($n$) and $\times4$ ($m$) for the 650-GHz receiver LO chains. The available power from the LO source is above 15~dBm, 12~dBm, 10~dBm, at 163~GHz, 220~GHz and 325~GHz, respectively.

 \begin{figure}[h!]
    \centering
    \includegraphics[width=0.9\linewidth]{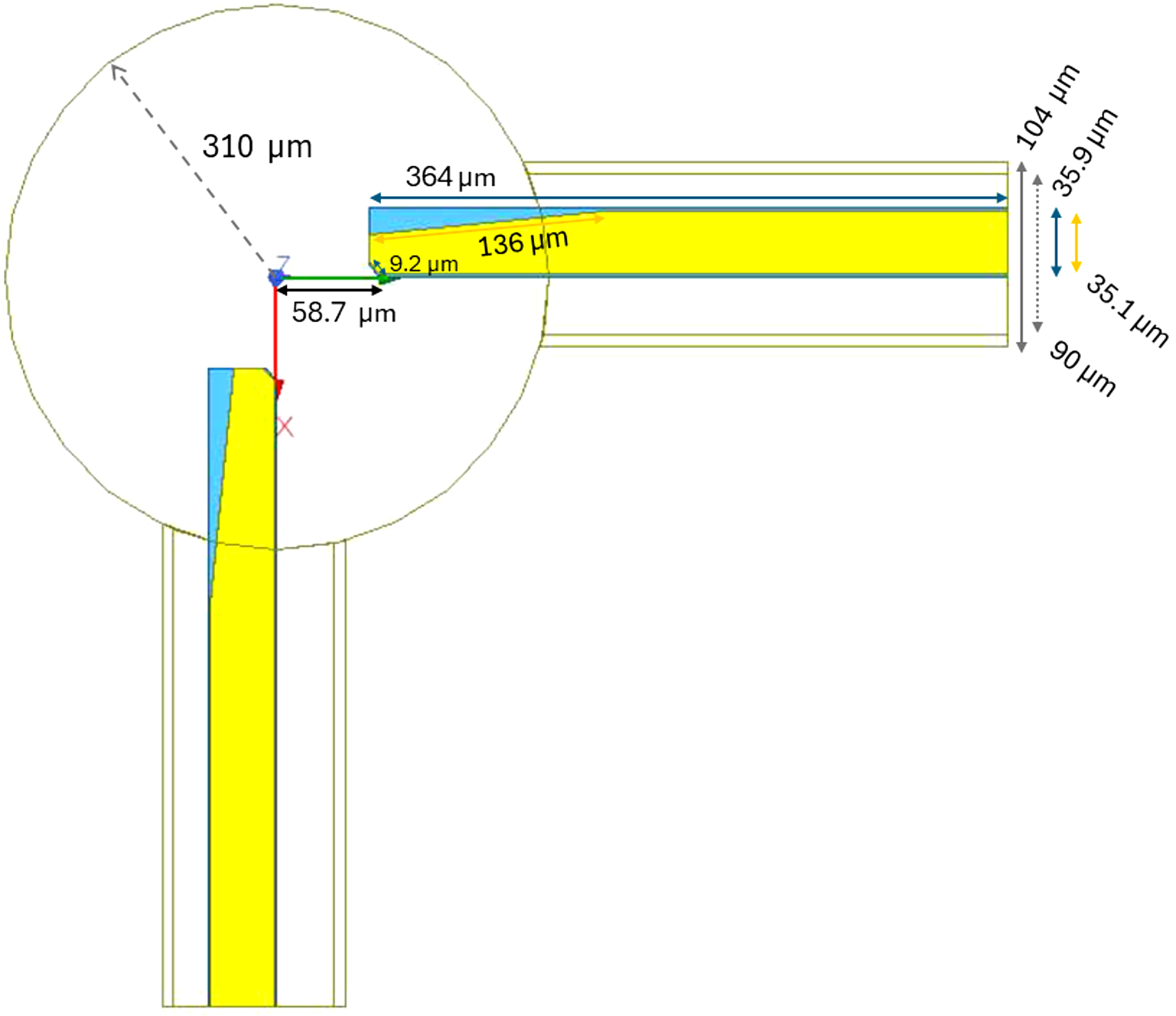}
    \caption{Asymmetric waveguide probes. Dimensioned drawing of the orthogonal-mode RF-probe transitions from a suspended stripline to a circular waveguide operating at 650~GHz.}
    \label{fig:OMTdim}
\end{figure}

\subsection{Conical Horn Antenna}
The design of the conical horn antenna is based on the smooth-walled spline profile presented by Granet \textit{et al.} in 2004 \cite{Granet_2004}. The spline coefficients were optimized to achieve high Gaussicity, the desired beam waist, and a positive derivative along the spline contour, thus enabling single-step drilling from the horn aperture to the circular waveguide interface with micron-scale precision \cite{Hammar_2016}.  These antennas have excellent performance due to their close integration with the mixer, which reduces waveguide losses.

The circular waveguide feed of the horn is directly machined as a single mechanical block, eliminating any additional waveguide interfaces and associated losses. Additionally, the horn antennas are designed to be fabricated using a single high-precision numerical-milling process, from the horn aperture at the top of the receiver to the circular waveguide section containing the RF probes. This method eliminates potential misalignment issues that can occur in conventional split-block horn assemblies \cite{Jayasankar2025}, which are especially critical in dual-polarization instruments. However, the 325-GHz horn antenna was manufactured as a conventional \textit{E}-plane split-block with the same profile as the antenna used on the Arctic Weather Satellite for testing purposes \cite{Albers2023}. The split is placed at \ang{45} of the orthogonal probes. Alignment features were added to ensure precise registration of the split planes. A misalignment between the top and bottom mechanical block may occur at the circular waveguide aperture where the RF probes are. Alignment features are added in the mechanical housing, in order to reduce misplacement down to a few microns.
The simulated antenna beam parameters for the three conical horns of spline profile operating at 325~GHz, 424~GHz and 650~GHz are summarized in Table~\ref{table:beamresults}. The antenna requirements were specific to different integration of the receivers into airborne or space borne radiometers. They can be tuned and optimised depending on the requirements of the mission.

\begin{table*}[t]
\begin{center}
\caption{Simulated horn antenna parameters.}
\label{table:beamresults}
\begin{tabular}{|c|c|c|c|c|c|c|c|}
\hline
\textbf{Frequency} & \textbf{WG diameter} & \textbf{Aperture diameter} &  \textbf{Horn length} & \textbf{Gaussicity} & \textbf{Gain} & \textbf{Return Loss} & \textbf{Beam Waist}\\ 
\textbf{(GHz)} &  \textbf{(mm)} &\textbf{(mm)} & \textbf{(mm)} & \textbf{(\%)} & \textbf{(dBi)} & \textbf{(dB)} & \textbf{(mm)} \\ 
\hline
325 &  0.62 & 7.5  &  23 & 97 &  25 &  30 & 2.0 \\ 
\hline
424 & 0.48 & 4.0 & 13 &  98 &  22 & 19 & 1.0 \\ 
\hline
650& 0.31 & 4.2 & 8.0 &  97 &  24 & 27 & 0.8 \\ 
\hline
\end{tabular}
\end{center}
\end{table*} 

\subsection{Orthogonal Mode, RF-Probe Transitions}
\begin{figure}[h!]
    \centering
    \includegraphics[width=0.9\linewidth]{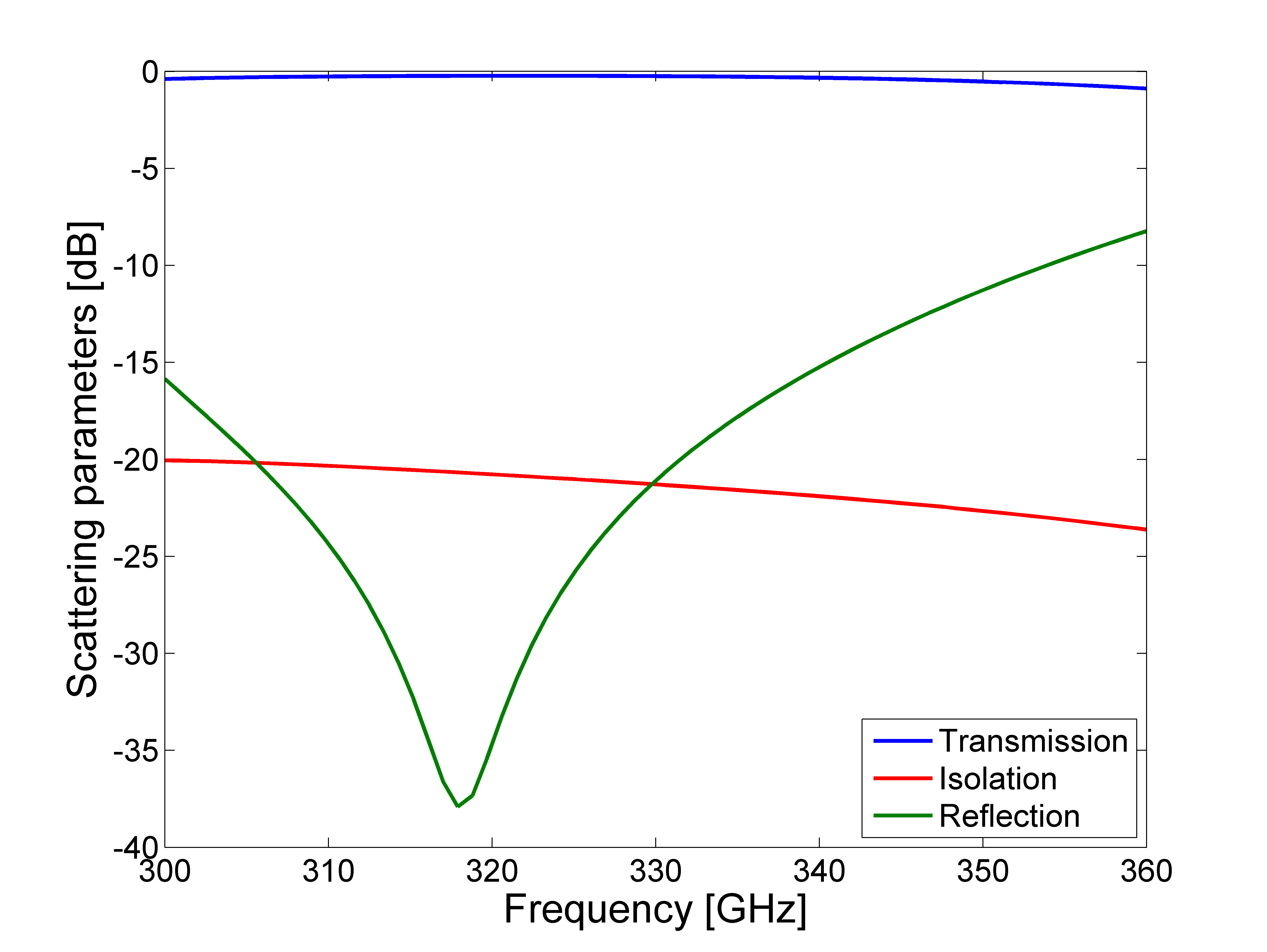}
    \caption{Planar orthomode transducer. Electromagnetic simulation of the performance of the 325-GHz orthogonal mode, RF-probe transition.}
    \label{fig:OMTresults}
\end{figure}

The two orthogonal degenerate polarization modes ($\mathrm{TE_{11}}$) in the circular waveguide are converted into suspended stripline modes using open-ended \textit{E}-field RF probes. The probes consist of a $1~\mu$m-thick gold stripline deposited on the same $3~\mu$m-thick gallium arsenide (GaAs) membrane and positioned asymmetrically within the circular waveguide, as shown in Fig.~\ref{fig:OMTdim}. Each probe is slightly offset and tapered to improve the cross-polarization isolation between the horizontal (H) and vertical (V) polarization channels \cite{Sobis2018}. A load imbalance between the two RF probes would generate a high input return loss, resulting in a reduce coupling efficiency, frequency-dependent ripples and higher insertion loss. However, the e-beam process with hundreds of nanometer accuracy to fabricate the microstrip and substrate of the RF probe almost remove such imbalance. Fig.~\ref{fig:OMTresults}  shows the electromagnetic simulation results of the dual probe transmission for the 325-GHz receiver. The model used includes the strip line length and cold circular waveguide length up to the antenna aperture. The simulation results indicate an input return better than 10 dB over 295-355~GHz, 380-460~GHz and 590-710~GHz (operational RF bandwidth). The isolation is found of 20-23~dB, 20-23~dB, and 21-24~dB, respectively over the operational bandwidth. Finally, the simulated insertion loss is 0.2-0.8~dB, 0.2-0.7~dB and 0.3-0.8~dB. The fractional bandwidth, the relative frequency range over which the antenna meets is specification (defined at 10 dB input return loss) is 18.5\% for the 325-GHz receiver, 19.0\% for the 424-GHz receiver and 18.5\% for the 650-GHz receiver.

\subsection{Heterodyne receiver} 
 The 325-GHz, 424-GHz, and 650-GHz subharmonic Schottky barrier diode mixer designs are based on the broadband 600-GHz mixer prototype delivered in 2014 to the Max Planck Institute for Solar System Research (MPS) as part of the front-end receiver development for the Submillimetre Wave Instrument (SWI) onboard the ESA JUpiter ICy moons Explorer (JUICE) mission \cite{Sobis2015}. The mixer design for the SWI mission will be detailed in an upcoming publication [Stake et al. in preparation]. The probe is integrated with the subharmonic diode mixer on a thin GaAs substrate, suspended between the waveguide \textit{E}-plane split, using freestanding gold beam leads that provide mechanical support, electrical ground, and IF-circuit connections \cite{Drakinskiy2013}. For each mixer, the simulated mixer conversion loss ranges from 8.7~dB to 10.1~dB for local oscillator (LO) input powers between 2~mW and 3~mW to efficiently pump the mixers.

An integrated 3-dB power splitter distributes the local oscillator (LO) signal from a single LO waveguide port. A matched Y-junction is integrated into the receiver housing, with the \textit{E}-plane split feeding the two mixers separately \cite{Kerr2001}.  

For each channel, a baseband (IF) low-noise amplifier MMIC based on an InP HEMT from Low Noise Factory AB (LNF-LNR1-15A) is integrated into the receiver module close to the mixer. The distance between the Schottky mixer and the low noise amplifier will determine the frequency and presence of IF ripples. To minimize the impacts of those ripples, a custom matching network is used. This matching network, biasing circuitry, and filter boards are integrated within the receiver, identically to the ones designed and assembled in the 183~GHz receiver described by Anderberg \textit{et al.} \cite{Anderberg2019}. The LNA offers a 1–15~GHz bandwidth (50~$\Omega$ system), a typical noise temperature of 54~K, and a gain of 37~dB at room temperature. 

\subsection{Fabrication and Receiver Assembly}

\begin{figure}[h!]
    \centering
    \includegraphics[width=\linewidth]{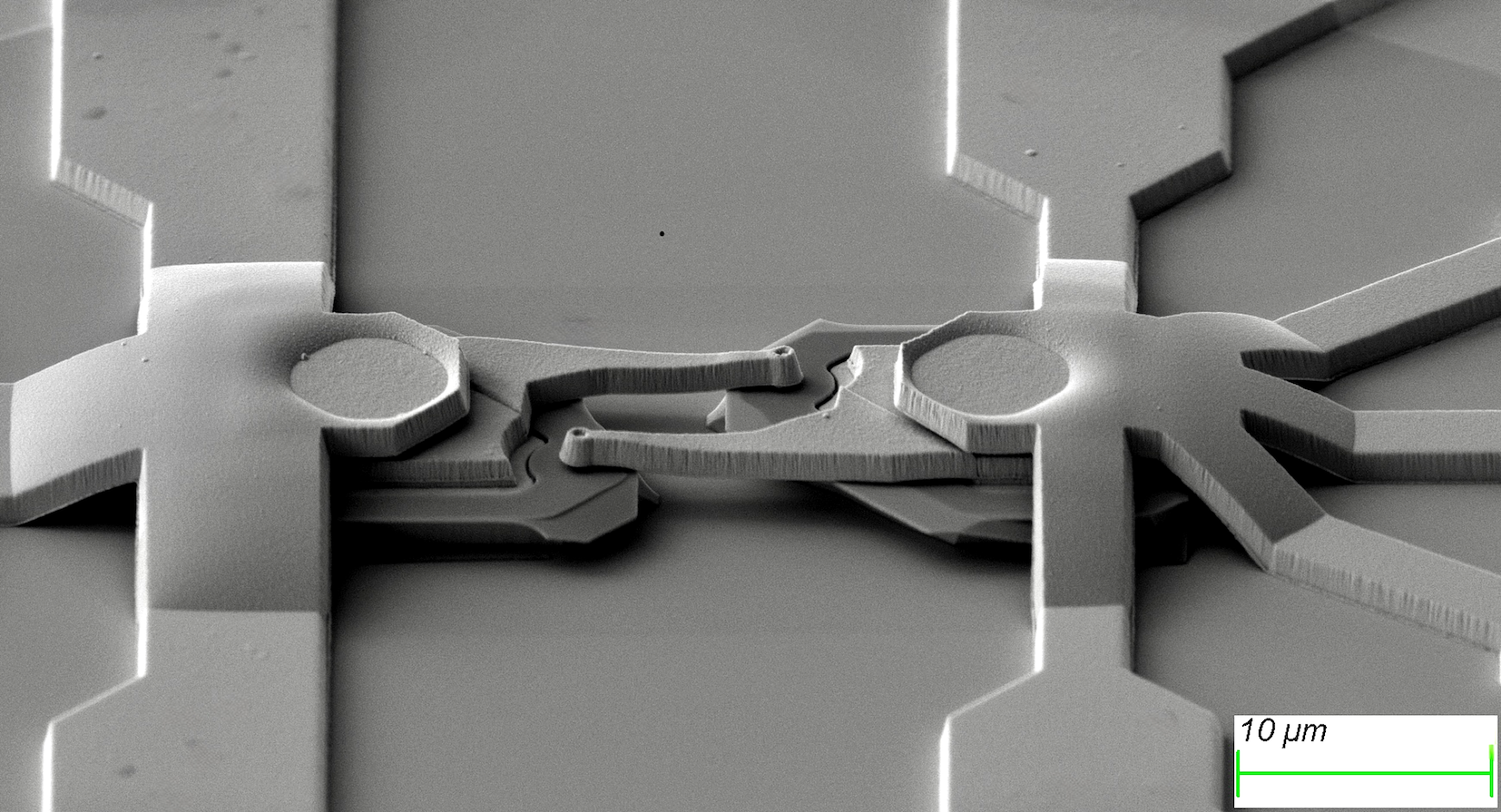}
    \caption{Integrated mixer circuit. SEM image of an anti-parallel, GaAs Schottky barrier diode pair part of the 650-GHz integrated mixer circuit, produced at the nanofabrication facility at Chalmers University of Technology.}
    \label{fig:didoe_sem}
\end{figure}

The integrated GaAs Schottky barrier diode circuits were fabricated at Chalmers University of Technology using electron-beam lithography, enabling precise, repeatable formation of anodes and air-bridges \cite{Drakinskiy2013}. The epitaxial structure consists of a 64-nm-thick cathode layer with an \textit{n}-doping concentration of $3\times 10^{17}$~cm$^{-3}$, grown on a 1.5-$\mu$m-thick highly-doped ($n^{++}$) contact layer. The device stack is supported on a 3-$\mu$m-thick undoped GaAs membrane. Fabrication starts by protecting the device layer with a stress-balanced $\mathrm{SiO_2}$ layer deposited using PECVD. Ohmic contacts are created through wet etching, metal deposition, lift-off, and annealing. Submicron anode contacts are then formed by wet etching through the $\mathrm{SiO_2}$ layer, depositing the Schottky metallization, and performing lift-off. After visual inspection and confirmation of anode size by scanning electron microscopy (SEM), air-bridge connections and diode isolation are completed.  The membrane is formed by selective wet etching of the 3-$\mu$m GaAs layer, which stops at the underlying (Al,Ga)As layer. Passive circuitry, such as beam leads, waveguide probes, and filter structures, is fabricated by depositing metallization and then performing lift-off. Finally, the sample is thinned from the backside to the (Al,Ga)As layer, which is then etched away to release the integrated diode circuits. On-wafer measurements of the diodes are conducted to characterize their I-V responses. The dc-series resistance and ideality factor for the mixer chips at 325~GHz, 424~GHz, and 650~GHz are 18–28~$\Omega$ and 1.18, respectively. SEM images of the anti-parallel Schottky barrier diodes are shown in Fig.~\ref{fig:didoe_sem}.

\begin{figure}[h!]
    \centering
    \includegraphics[width=\linewidth]{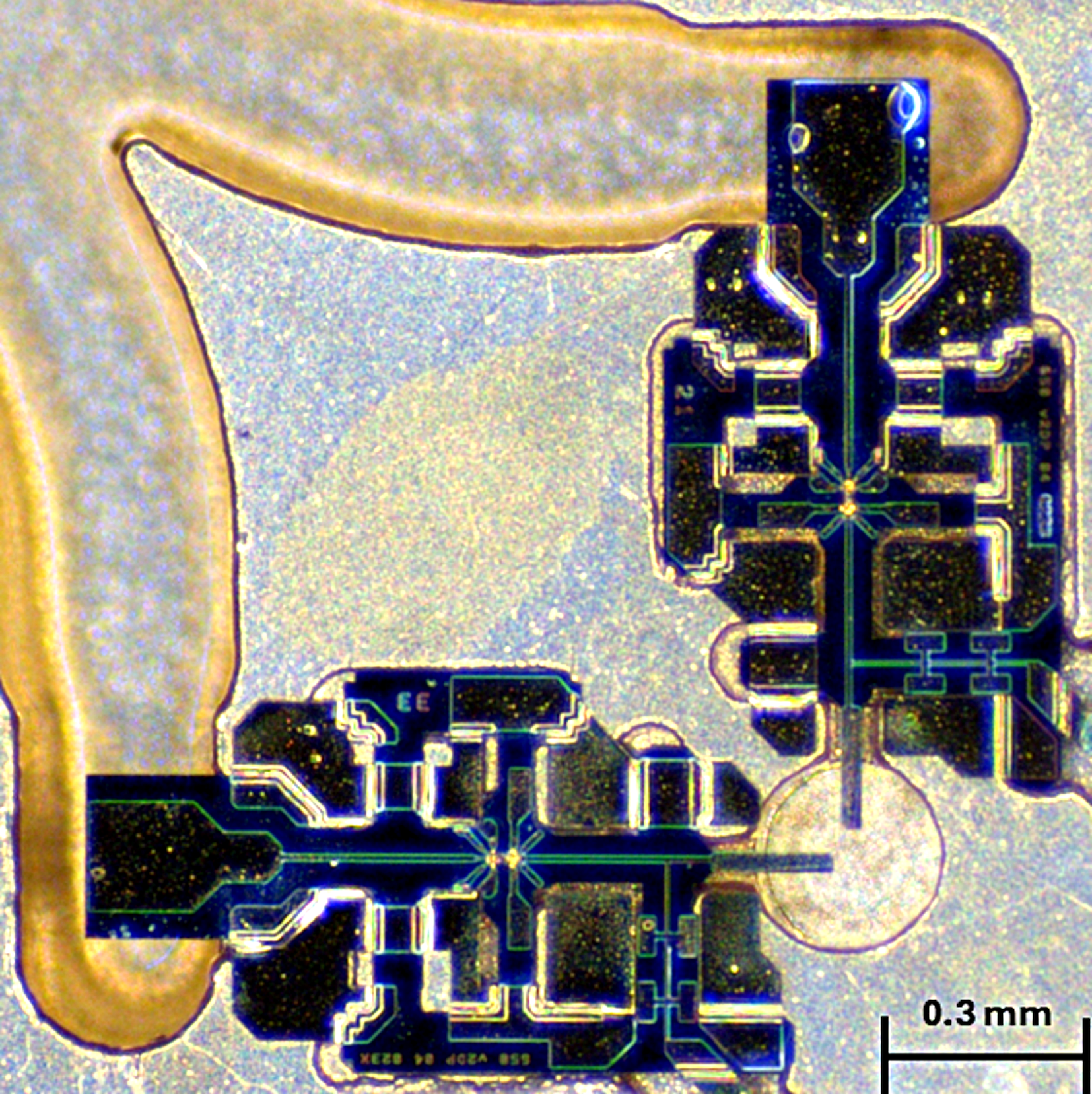}
 \caption{Mixer circuits assembly. Top view of the mounted 650~GHz dual-polarization integrated mixers, with LO waveguide splitter ports and antenna circular waveguide.}
    \label{fig:receivermounta}
  \end{figure}

\begin{figure}
    \centering
    \includegraphics[width=\linewidth]{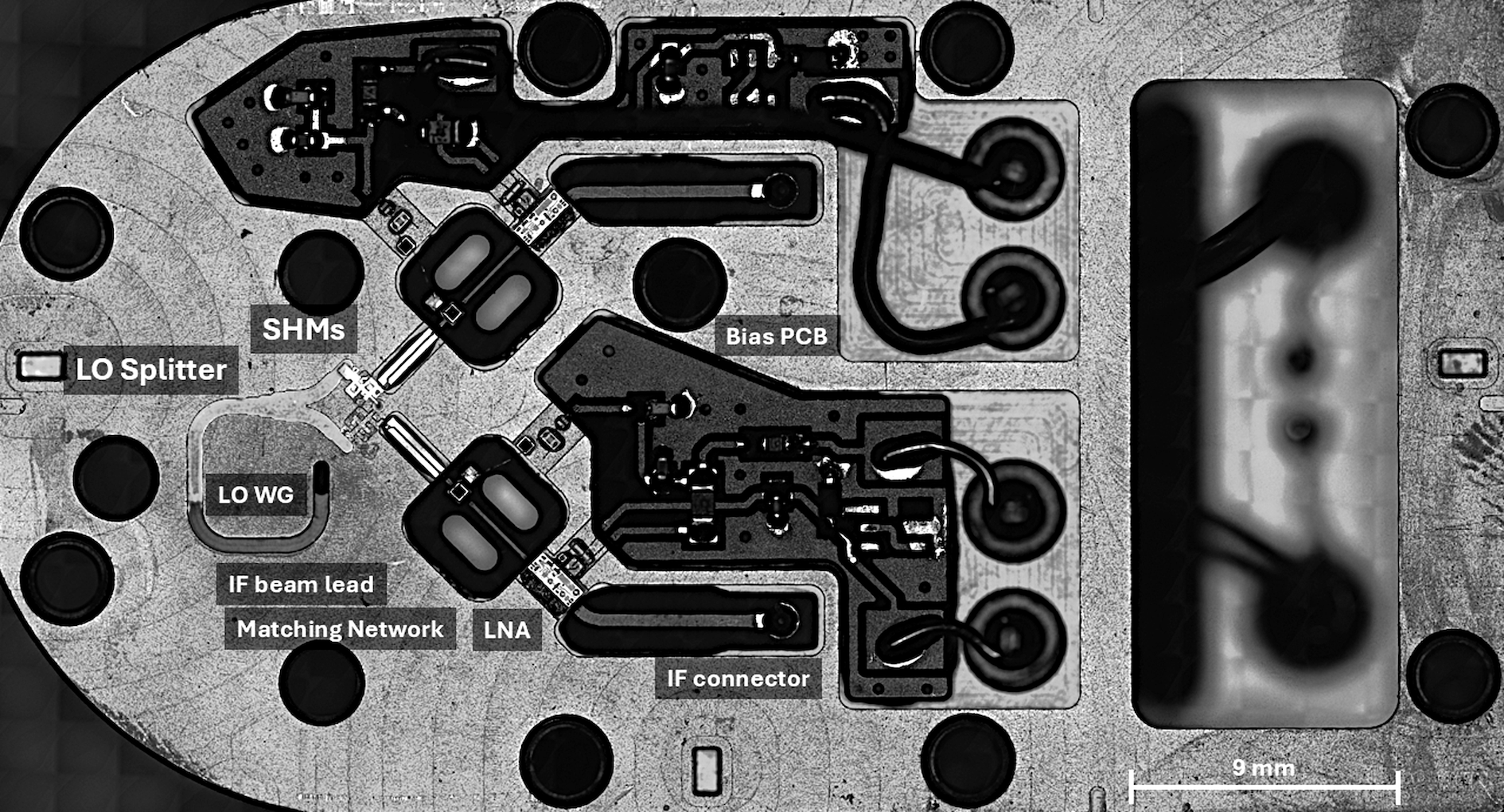}
    \caption{Receiver module assembly. Picture of the 424-GHz dual-polarization receiver during assembly, with one integrated mixer IC, LO waveguide ports, IF matching network, and biasing PCBs.}
    \label{fig:receivermountb}
  \end{figure}

\begin{figure}
    \centering
    \includegraphics[width=\linewidth]{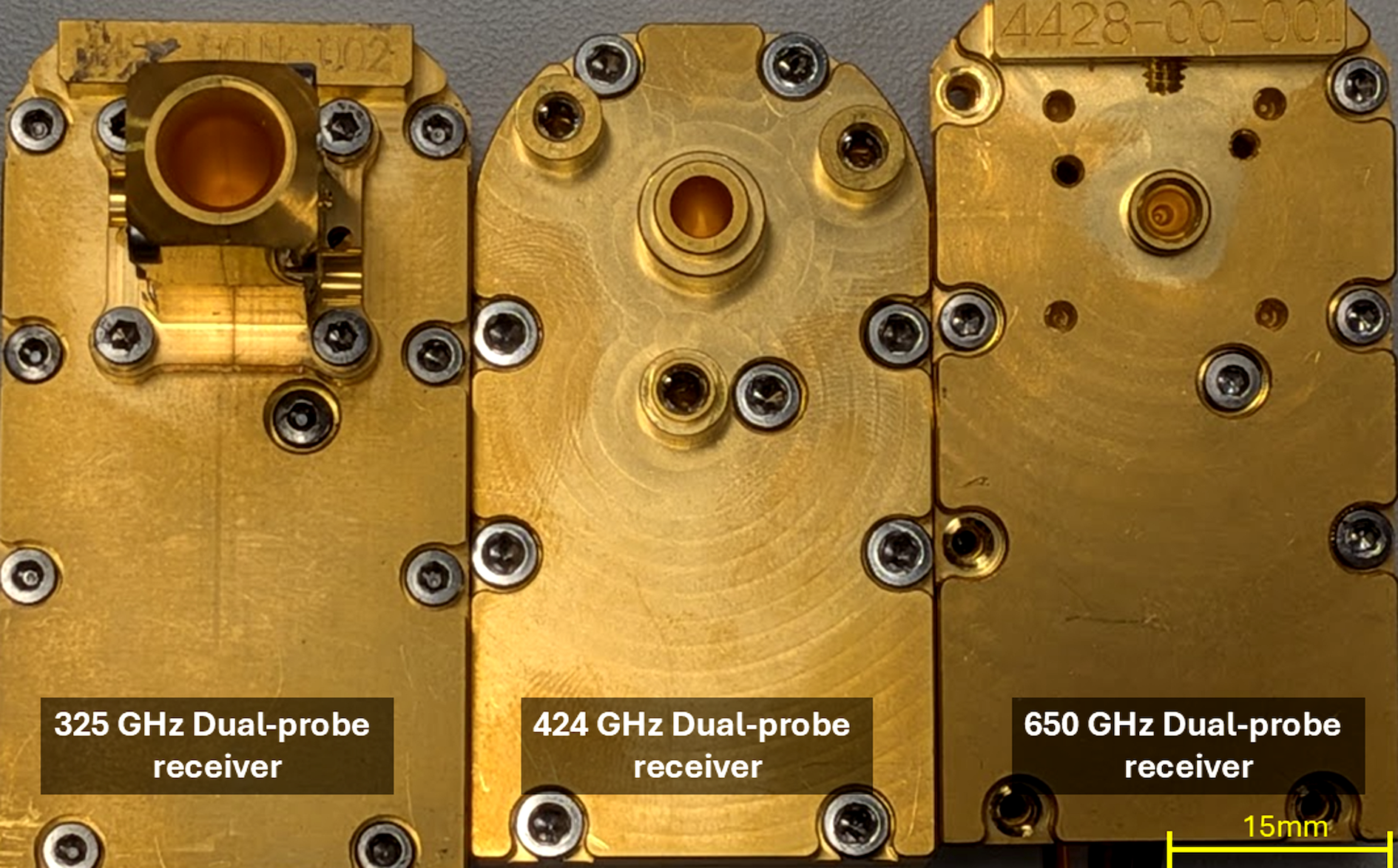}
    \caption{Dual-polarization receiver modules. Top view of assembled receivers operating in the 325~GHz, 424~GHz, and 650~GHz bands.}
    \label{fig:receivermountc}
    \end{figure}

The receiver waveguide housing was manufactured from aluminum using a high-precision CNC milling machine. Alignment fixtures ensure the correct positioning of the horn and LO waveguide interfaces, including both outer and inner interfaces. All parts were measured before gold plating. The accuracy of the integrated horn profile is verified by comparing a section of the horn cut in half to the theoretical spline profile. For the 650~GHz split-half horn, the maximum manufacturing deviation from the theoretical shape is $14 \pm 6~\mu$m, while for the integrated horn section, it is $19 \pm 12~\mu$m. Optical inspection confirms the shape and surface quality of the fabricated horn apertures. The receiver blocks were assembled at the AAC Omnisys facility. The two mixer circuits were placed orthogonally within the mechanical housing, as shown in Fig.~\ref{fig:receivermounta} and Fig.~\ref{fig:receivermountb}. Fig.~\ref{fig:receivermountc} shows the top mechanical housing of the receiver modules with the integrated horn. 

The \textit{E}-probes will fit within the horn circular waveguide, and the two mechanical block halves will permanently fix the two integrated mixer circuits. Misalignment may arise between the two \ang{90} mixer circuits due to assembly difficulties or probe placement within the circular waveguide, thereby degrading the cross-polarization isolation. 

\subsection{Receiver Characterization}\label{subsectseup}

\begin{figure*}
  \centering
  \begin{subfigure}[b]{0.45\textwidth}
    \centering
    \includegraphics[width=\linewidth]{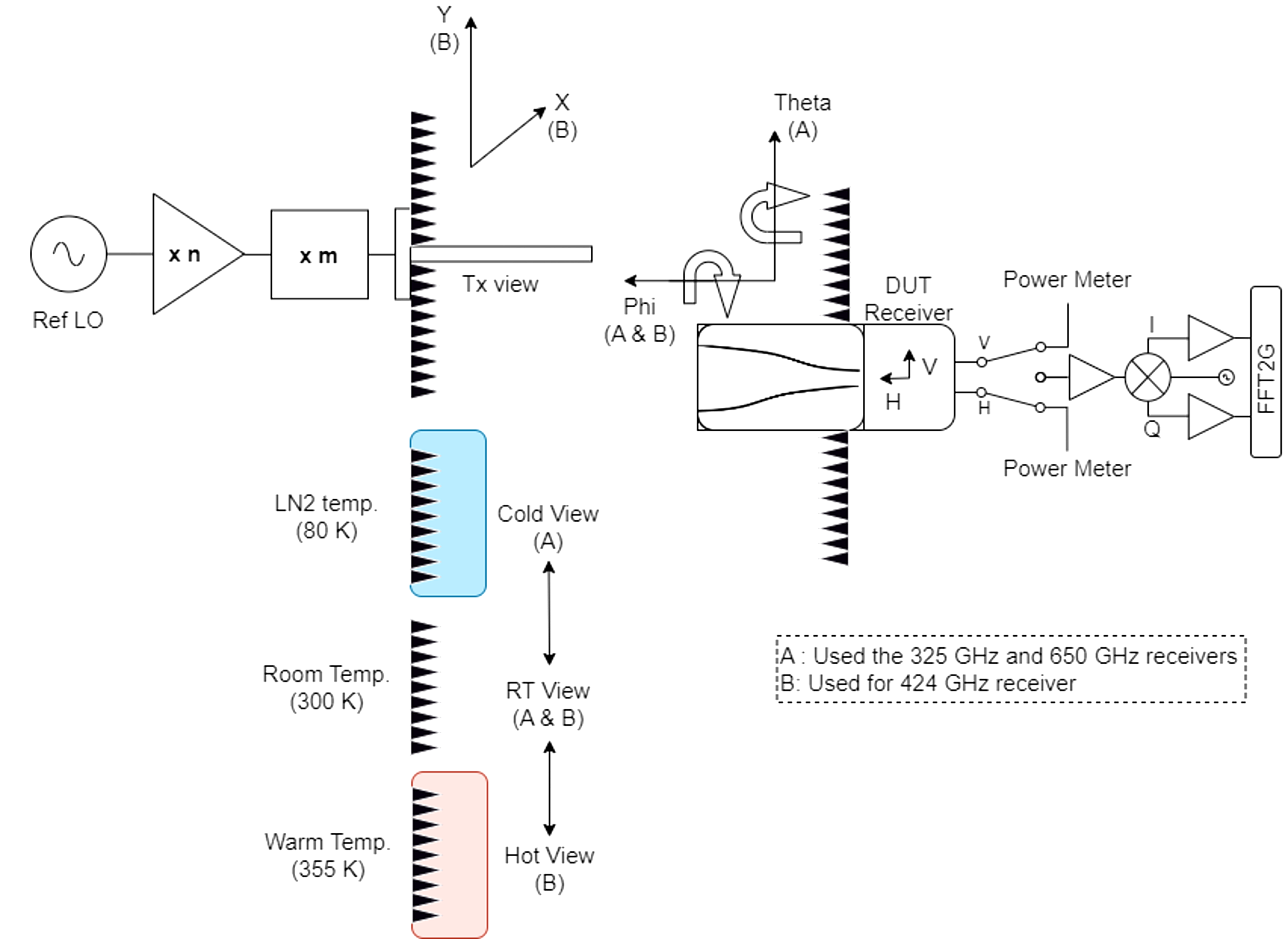}
  \end{subfigure}
  \begin{subfigure}[b]{0.45\textwidth}
    \centering
    \includegraphics[width=\linewidth]{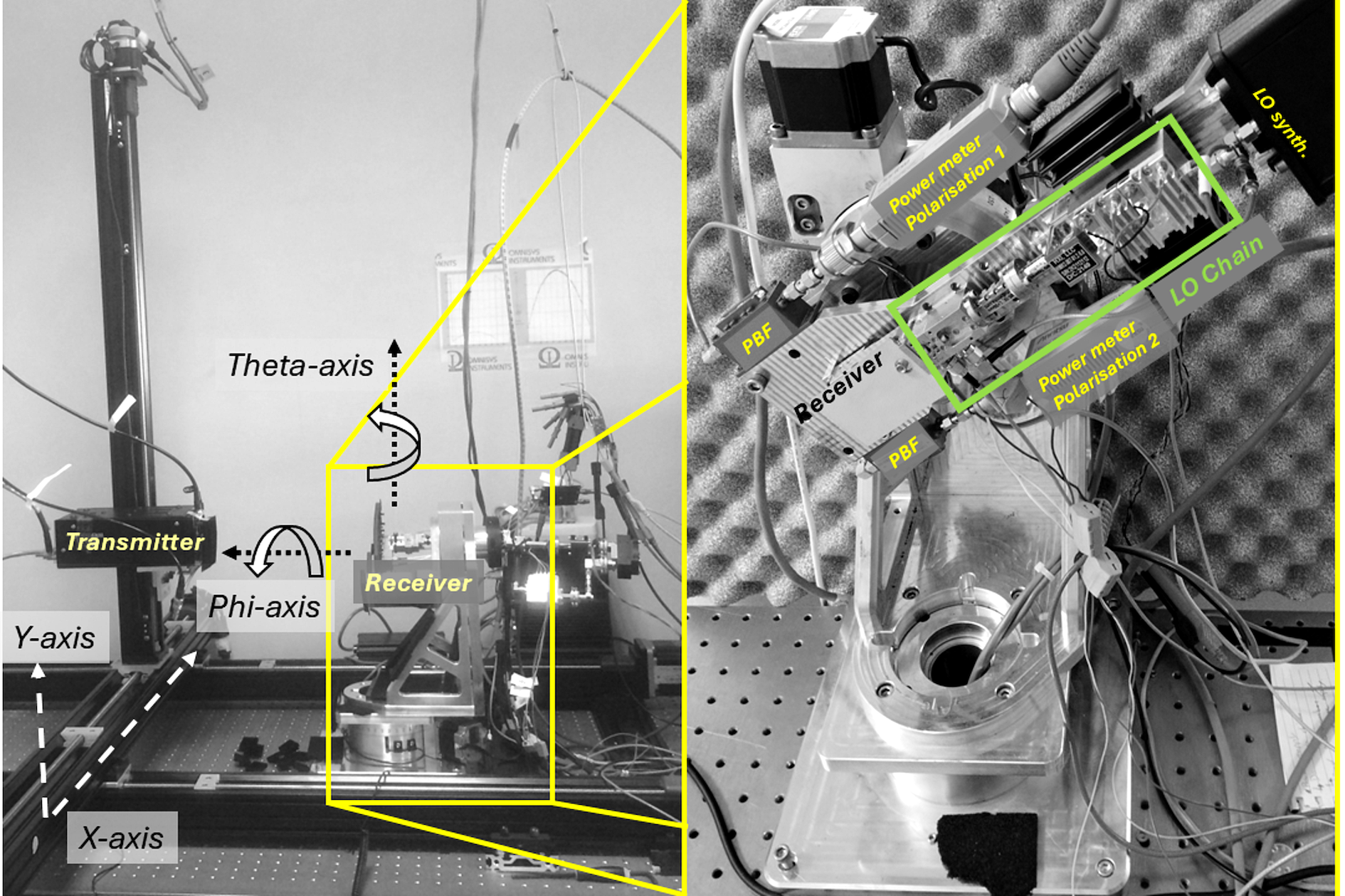}
  \end{subfigure}
  \caption{(Left) Schematic of the radiometric and antenna test set-ups, showing the transmitter, the cold blackbody target, and the room temperature blackbody target. (Right) Picture of the quasi-optical antenna test set-up, showing the XY-stage, the $\phi$ and $\theta$ stage, the transmitter, and the receiver. }
    \label{fig:set_up}
\end{figure*}

The double-sideband (DSB) receiver noise temperature was measured using the Y-factor technique, while the cross-polarization performance of the dual-polarization concept and the antenna beam patterns were characterized using a far-field setup with rotating stages around a static transmitter, as shown in Fig.~\ref{fig:set_up}.

Noise measurements of the receiver system were obtained by recording the output power from a room temperature blackbody (300~K), a hot calibration target (355~K), and/or a liquid-nitrogen-cooled blackbody (80~K). The 325-GHz and 650-GHz receivers were measured using room-temperature and liquid-nitrogen targets. In contrast, the 424~GHz receiver was characterized using room-temperature and hot targets, as described by Hammar \textit{et al.} \cite{Hammar2018}.  The receiver’s IF outputs are connected to an IQ mixer and amplifier stage followed by AAC Omnisys' 5~GHz fast-Fourier-transform (FFT) digital spectrometer, which provides a spectral resolution of 0.3052~MHz across 16,384 channels \cite{FFT2G}. The IF measurement bandwidth spans 0.5–18~GHz with an integration time of 1~s.

The far-field antenna beam measurement setup comprises two main components (as shown in Fig.~\ref{fig:set_up}): a XY 2-axis linear stage for the RF source, which transmits a tone via an open waveguide, and a dual-rotational stage that holds the dual-polarization receiver to acquire the $\phi$ (polarization) and $\theta$ (beam pattern) angles. The first rotation axis is centered on the integrated feedhorn aperture, allowing rotation of the receiver polarizations (H or V) about the transmitter axis over a range of $\pm 60^\circ$. The second axis is centered on the receiver feedhorn beam waist, allowing rotation of the horn relative to the source over $\pm 40^\circ$. Blackbody absorbers are placed at both receiver and transmitter RF ports to suppress reflections.  

For the 424~GHz receiver, the co-polar and cross-polar antenna patterns were measured by scanning the RF open-waveguide transmitter source in the XY-plane in front of the receiver and rotating it by 90$^\circ$ for the cross-polar radiation pattern. It was not possible to use the $\phi$ and $\theta$ rotating stage of the receiver at the time of the 424~GHz receiver measurements. The 424~GHz receiver results have a higher measurement uncertainty of 2.5~dB, due to the experimental setup. 

For the 325~GHz and 650~GHz receivers, the RF open-waveguide transmitter source was kept aligned with the center of the horn aperture, and the $\phi$ and $\theta$ rotating stages were used to measure the cross-polarization isolation and E-plane co-polar radiation patterns.
To reduce the measurement noise floor, bandpass filters (4~GHz center frequency, 6~MHz bandwidth, Mini-Circuits ZVBP-4000-S+) are connected at the IF ports, followed by a high-sensitivity power sensor (Anritsu ML 2438A). The overall dynamic range of the beam measurement system is approximately 40~dB, and the measurement uncertainty for the cross-polarization measurements is 0.4~dB.

The frequency multiplication coefficients for the transmitter chain $n$ and $m$ in Fig.~\ref{fig:set_up} correspond to $\times8$ ($n$) and $\times
4$ ($m$) for the 325~GHz receiver, $\times36$ ($n~\&~m$) for the 424~GHz receiver, and $\times8$ ($n$) and $\times8$ ($m$) for the 650~GHz receiver LO chains.

\section{Results}
\subsection{Far-field Antenna beam Patterns and Cross-Polarization Isolation}

\begin{figure}[h]
    \centering
    \includegraphics[width=1\linewidth]{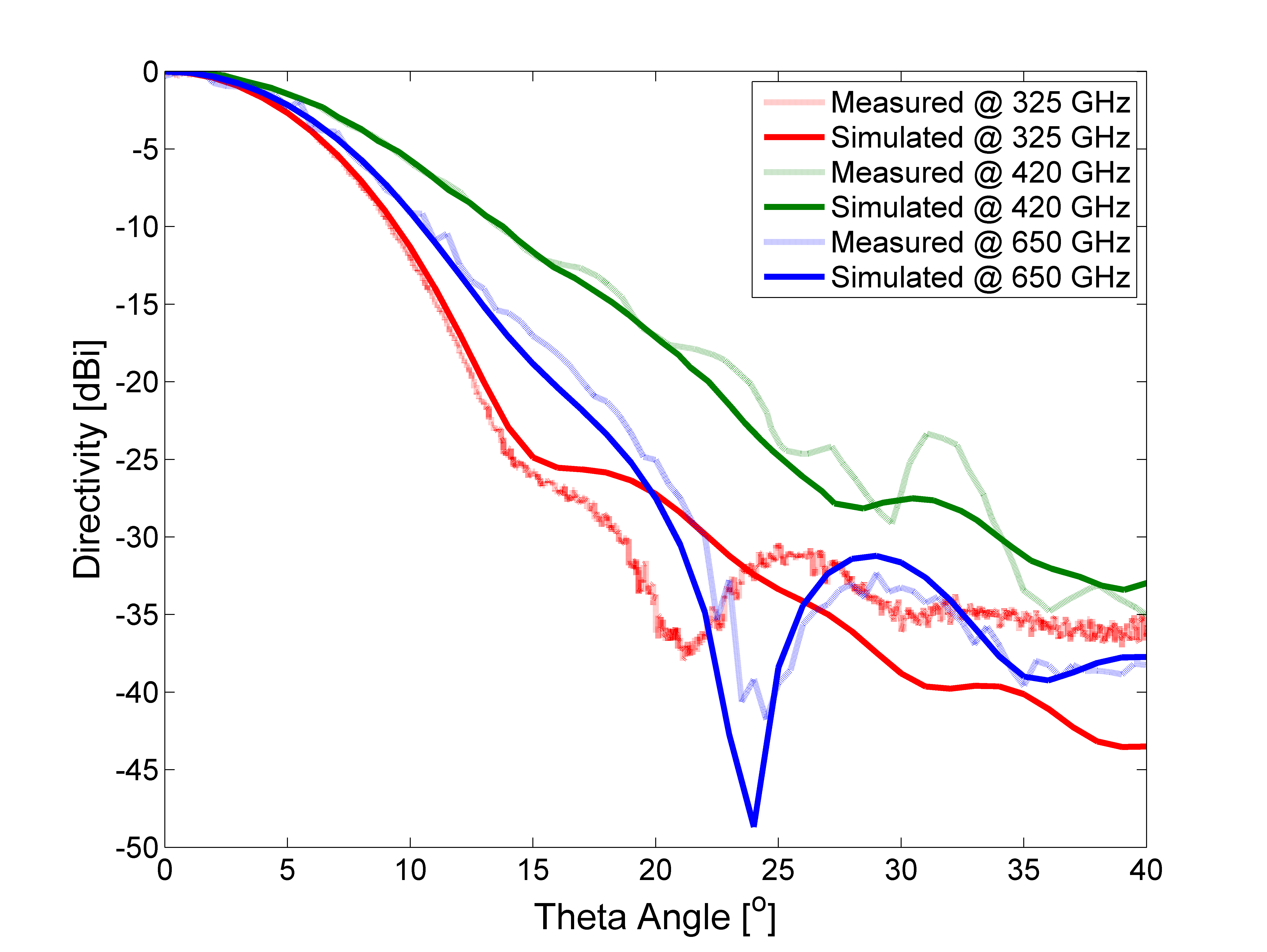}
    \caption{Antenna directivity. Normalized far-field \textit{E}-plane radiation patterns of the three spline horn antennas measured at 325~GHz (red), 424~GHz (green), and 650~GHz (blue).}
    \label{fig:beamresults}
\end{figure}

The measured far-field antenna beam patterns for the three receivers, compared with simulations, are shown in Fig.~\ref{fig:beamresults}. The full electromagnetic simulation and measured beam pattern show good agreement, consistent with the manufacturing precision analysis.

\begin{figure}[h]
    \centering
    \includegraphics[width=1\linewidth]{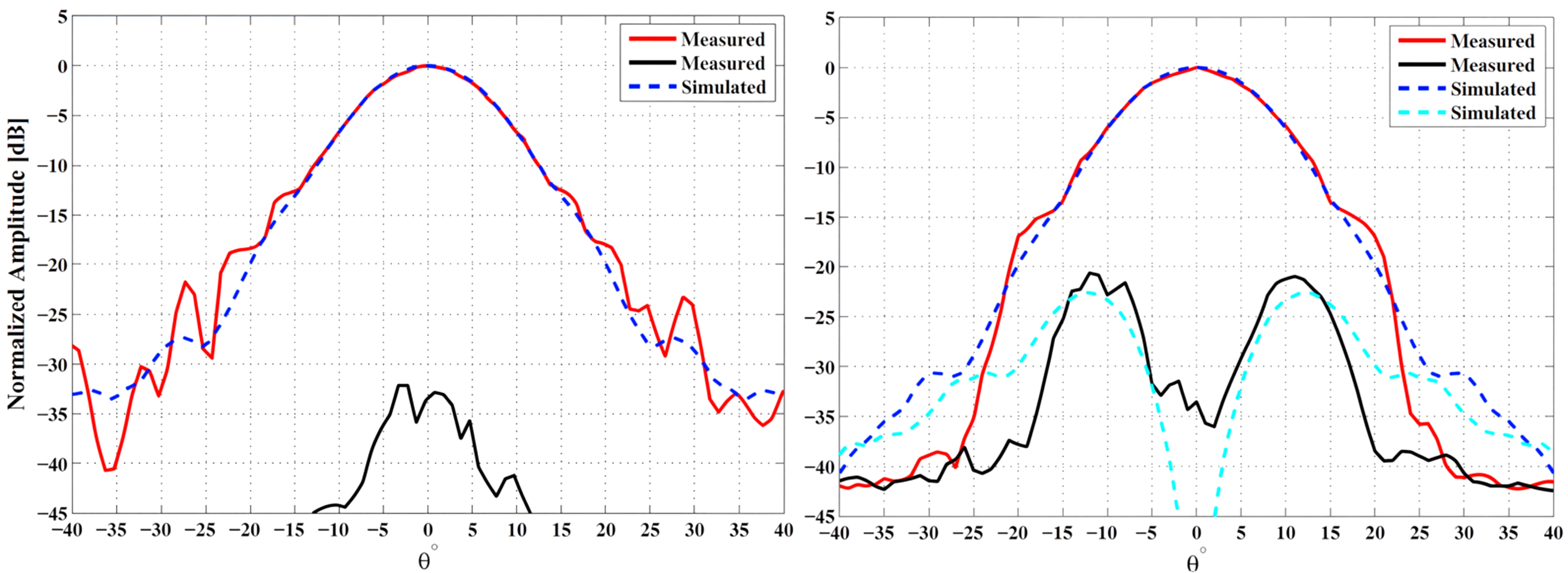}
    \caption{Antenna directivity. Normalized far-field \textit{E}-plane and \textit{D}-plane radiation patterns of the 424-receiver spline horn antenna, measured at 420.2~GHz.}
    \label{fig:DP2424_antenna}
\end{figure}

The E-plane and D-plane for the H-polarisation of the 424-receiver, with a LO set at 420.2~GHz are shown in Fig.~\ref{fig:DP2424_antenna}. The cross polar component for the E-plane could not be simulated due to symmetry planes that were required for a reasonable simulation time.

\begin{figure} [h]
    \centering
    \includegraphics[width=1\linewidth]{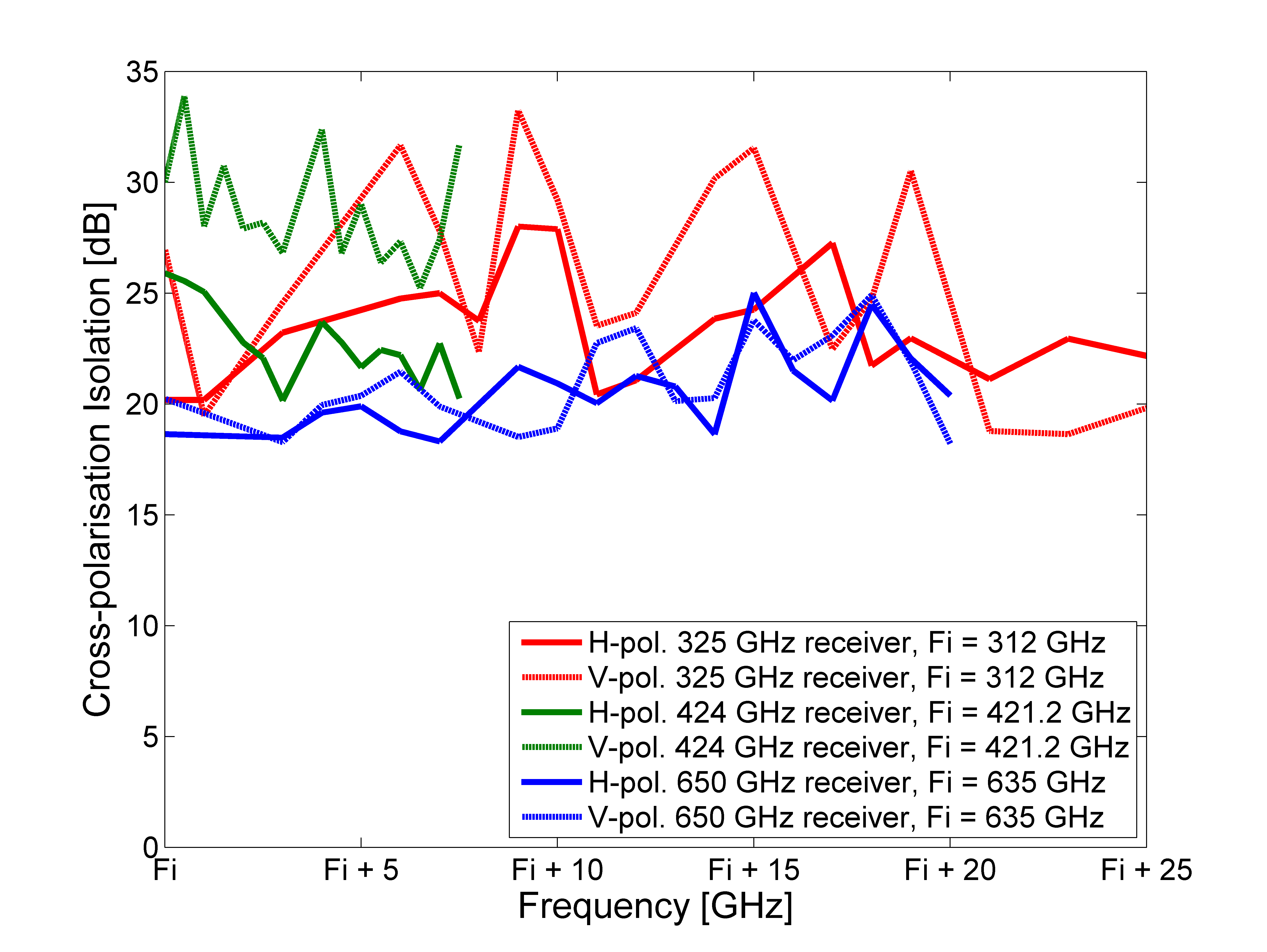}
    \caption{Cross-polarization isolation. Measured cross-polarization versus frequency range of the 325-, 424- and 650-GHz dual-polarization receivers.}
    \label{fig:crosspolresults}
\end{figure}

The cross-polarization isolation (XPI) \cite{Watson1973} between the H and V channels was measured at a $\theta=0^\circ$ angle relative to the transmitter by scanning the polarization angle $\phi$. For each $\phi$ angle, the H and V directivities were subtracted to obtain the receiver cross-polarization isolation parameter. Representative results over the frequency range of the receivers are shown in Fig.~\ref{fig:crosspolresults} with Fi being the lowest RF frequency measured. A good agreement with the simulated cross-polarization values is observed, with values ranging from $-19$~dB to $-34$~dB over the 312–655~GHz frequency range. Minor deviations between the measured cross-polarization isolation arise from residual standing waves in the experimental setup, amplitude imbalance between the two mixers or misalignment of the two dual-polarization RF probes relative to the antenna circular waveguide, which can reduce the orthogonal channel isolation. The potentially amplitude imbalance between the mixer ICs and IF LNA’s gain is compensated by the high dynamic range of the test- setup. The assembly misalignment error between the position of the H- and V-MMICs inside the receiver packaging would cause a decrease of the cross-polarization isolation by up to 2~dB for a \ang{10} phi angle error.

Table~\ref{resultsIXPcompa} summarizes the state-of-the-art results for orthomode transducers (OMTs) with the presented dual \textit{E}-plane probe transitions. The presented concept shows outstanding performance, combining high XPI across polarizations and low insertion loss at high frequencies. The planar orthomode RF probe transitions show low insertion loss of approximately 0.3 dB (simulated). 

\begin{table}[h]
\centering
\caption{Comparison of orthomode transducers characteristics operating above 200~GHz.}
\label{resultsIXPcompa}
\begin{adjustbox}{max width=0.5\textwidth}
\begin{tabular}{| c | c | c | c | c |}
\hline
Frequency & Insertion Loss & XPI & Technology & Reference\\
(GHz) & (dB) & (dB) &   &  \\
\hline
220 - 330 & 0.25 - 0.75 & $\geq$ 25 & Silicon Turnstile & \cite{GomezTorrent2019} \\
\hline
250 - 370 & 0.3 - 0.7 & $\geq$ 24 & CNC Machining & \cite{Lapkin2025} \\
\hline
280 - 330 & 3.0 & $\geq$ 23 & CNC Machining &  \cite{Stiewe2023}\\
\hline
300 - 400 & $\leqslant$ 1.0 & $\geq$ 20 & Probes & \cite{KuanYuLiu2013} \\
\hline

\textbf{312 - 338} & $\sim$ \textbf{0.3} * &  $\geq$ \textbf{19} & \textbf{Integrated dual probe receiver} & \textbf{This work} \\
\hline

300 - 500 & 1.5 - 0.5 & $\geq$ 35 & CNC Machining & \cite{Gonzalez2021} \\
\hline

\textbf{420 - 430} & $\sim$ \textbf{0.3} * & $\geq$ \textbf{20} & \textbf{Integrated dual probe receiver} & \textbf{This work} \\
\hline

500 - 600 & 0.6 & $\geq$ 25 & Silicon Machining & \cite{JungKubiak2016} \\
\hline
500 - 600 & 3.0 - 5.0 & $\geq$ 35 & CNC Machining &  \cite{Reck2013} \\
\hline
\textbf{635 - 655} & $\sim$ \textbf{0.3} * & $\geq$  \textbf{18} & \textbf{Integrated dual probe receiver} & \textbf{This work} \\
\hline

\end{tabular}
\end{adjustbox}
\caption*{* \scriptsize EM Simulations.}
\end{table}

\subsection{Receiver Stability and Noise Temperature}

\begin{figure} [h]
    \centering
    \includegraphics[width=1\linewidth]{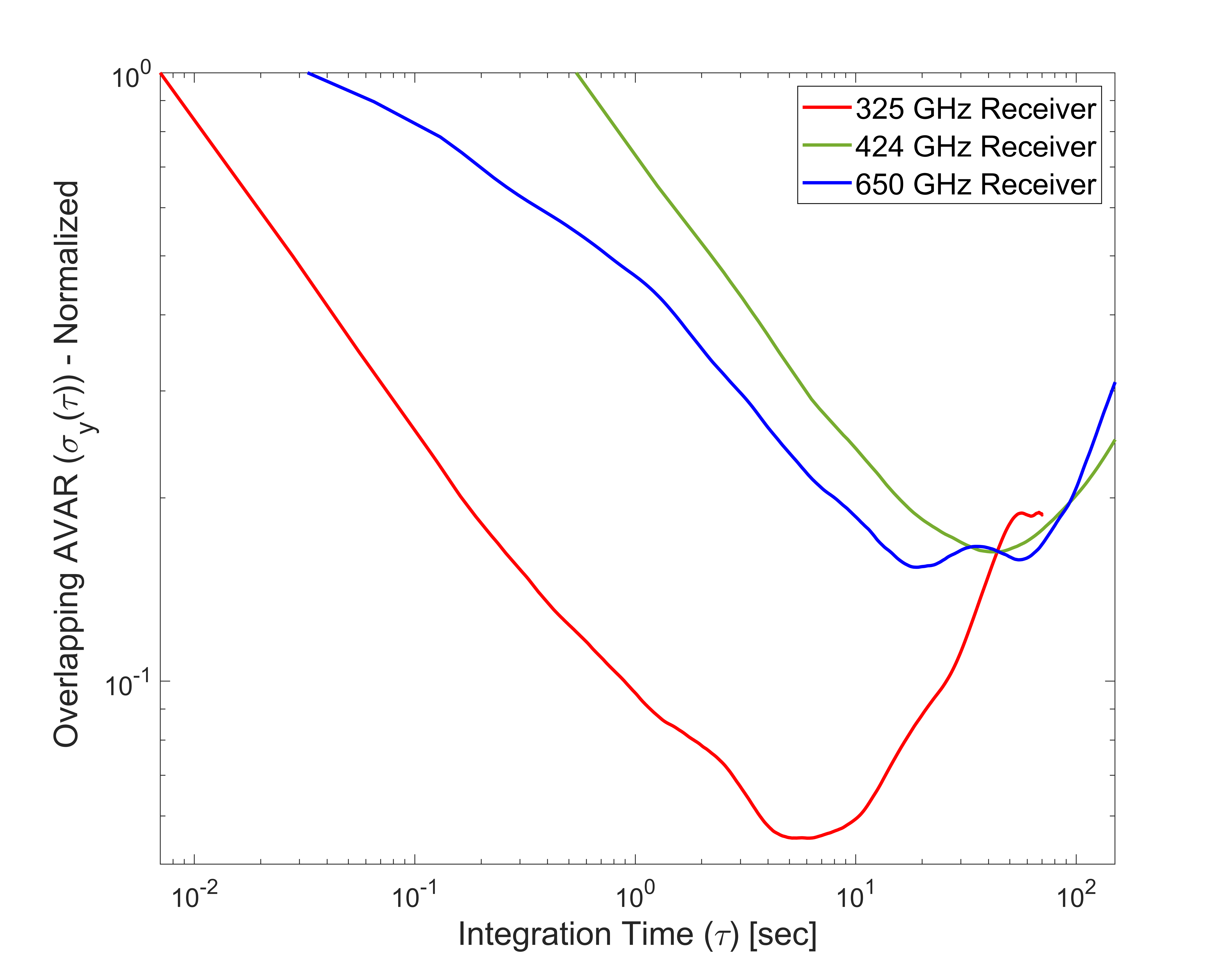}
    \caption{Receiver Stability. Overlapping Allan variance versus integration time (s) for the 325-GHz (red), 424-GHz (green), and 650-GHz (blue) receivers.}
    \label{fig:avarplot}
\end{figure}

The receiver stability was characterized using the Allan variance method \cite{Allan1966, Schieder2001}. Each receiver was placed in front of a constant-temperature calibration target at room temperature within a controlled laboratory environment. The calibration targets provided stable and uniform input noise power. The IF output power was recorded with a power sensor (Anritsu ML 2438A) for 5~minutes with a sample acquisition rate of 143~Hz for the 325-receiver, 6~Hz  for the 424-receiver and 30~Hz  for the 650-receiver because of different power meter settings at the time of the measurements. The Allan variance integration times, corresponding to the minimum noise contribution, were measured as 7~s, 12~s, and 11~s for the 325~GHz, 424~GHz, and 650~GHz receivers, respectively as shown in Fig.~\ref{fig:avarplot}. The variations between the receivers are due to different noise temperature, different bandwidth and minimum acquisition time used at the time of the measurements.

\begin{figure}[h]
    \centering
    \includegraphics[width=1\linewidth]{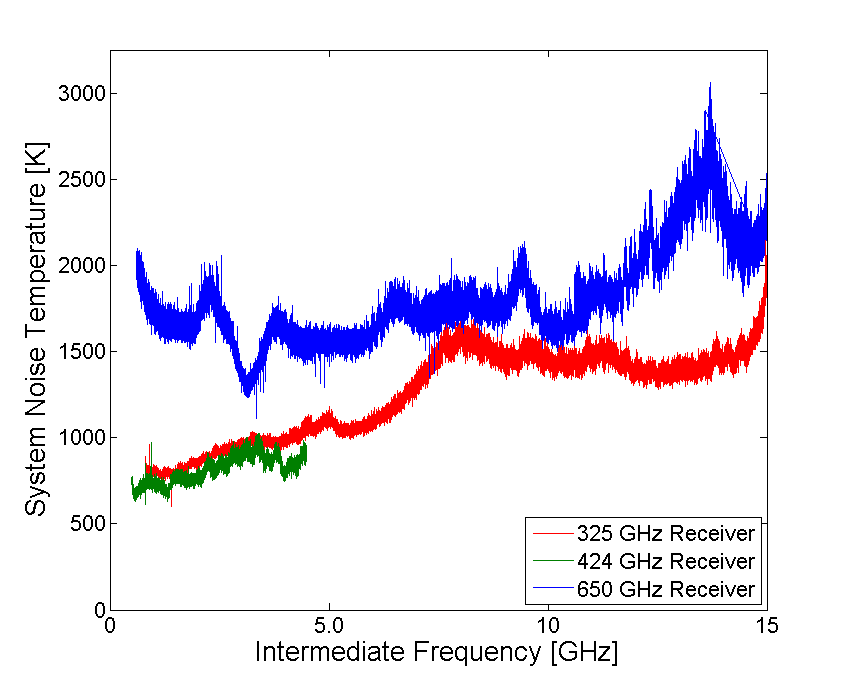}
    \caption{System noise temperature. DSB receiver noise temperature versus intermediate frequency (IF) for the 325-GHz (red), 424-GHz (green), and 650-GHz (blue) receivers with LO at 325~GHz, 421.7~GHz and 630~GHz, respectively.}
    \label{fig:trecplot}
\end{figure}

The double sideband (DSB) receiver noise temperature was measured using the Y-factor method, as described in Section~\ref{subsectseup}. After setting the LO frequency and power at the receiver input, the noise temperature was characterized across the full IF bandwidth of each receiver. The typical results are shown in Fig.~\ref{fig:trecplot} with LO at 325~GHz, 421.7~GHz and 630~GHz. The 424~GHz receiver is equipped with an IF LNA covering up to 4.7~GHz, whereas the 325~GHz and 650~GHz receivers incorporate wideband LNAs with bandwidths up to 15~GHz. All receivers exhibit their lowest noise temperatures around 2~GHz IF: below 1000~K for the 325~GHz and 424~GHz receivers, and 1800~K for the 650~GHz receiver. At IF frequencies above 8~GHz, poor matching between the mixer and LNA increases the noise contribution. The 325 GHz receiver system noise temperature appears higher than the 424 GHz receiver most likely because of either slight misplacement of mixer ICs in the receiver block or misalignment of the top block and horn antenna. This would slightly increase the insertion loss ahead of the mixer ICs.

A summary of the overall receiver noise temperatures, including contributions from the horn and probe transitions, is presented in Table~\ref{Trecresultssumary}. The average noise temperatures and standard deviations over the measured IF ranges were recorded for LO frequencies of 315~GHz and 325~GHz (325-GHz receiver), 421.7~GHz and 427.7~GHz (424-GHz receiver), and 630~GHz and 650~GHz (650-GHz receiver). The 325~GHz receiver exhibits an overall DSB noise temperature below 1200~K, with the lowest IF band below 900~K. The 424~GHz receiver maintains a noise temperature below 1000~K across its IF range. The 650~GHz receiver shows a DSB noise temperature below 1900~K at the lowest IF, ranging from 1600~K to 2300~K depending on the LO frequency and channel. The minimum measured DSB noise temperatures were 700~K, 680~K, and 1250~K for the 325-GHz, 424-GHz, and 650-GHz receivers, respectively. Each mixer from the V- and H-polarization branches may require slightly different LO power to achieve their highest RF performance. It is not possible to individually tune each mixer in the current dual-probe receiver configuration as they are unbiased and because of the integrated 3-dB power splitter. A future improvement on the concept would separate the LO splitter outside of the receiver mechanical block.
For comparison, these results are better than state-of-the-art single-polarization receivers at similar frequencies, such as those onboard the ICI mission 1200~K for the 325~GHz channel, 1400~K for the 448~GHz channel and 1800~K for the 664~GHz channel \cite{Thomas2023}. A slight tuning of the receiver frequency response of the presented receiver would permit to reduce drastically the noise equivalent differential temperature of the ICI mission. For example, by using the minimum mean noise temperature from Table~\ref{Trecresultssumary}), the presented receiver's in the ICI radiometer would improve the instrument sensitivity by 33~\% at 325~GHz, 43~\% at 448~GHz and 11~\% at 650~GHz.

\begin{table}[h]
\centering
\caption{Measured Receiver noise temperature ($T_\mathrm{sys}$)}
\label{Trecresultssumary}
\begin{tabular}{|c|c|c|c|}
\hline
LO freq. & IF Bandwidth & Mean H-pol. & Mean V-pol. \\
 & & {\scriptsize \textit{[Min., Max., Std.]}}  & {\scriptsize \textit{[Min., Max., Std.]}}  \\
(GHz) & (GHz) & (K) & (K) \\
\hline 
315 & 0.8 -- 8.0  & \textbf{850} & \textbf{800} \\
& &  {\scriptsize \textit{670, 1100, 120}} &  {\scriptsize \textit{700, 1100, 80}} \\
\hdashline
315 & 0.8 -- 15.0 & \textbf{1000} & \textbf{1100} \\
& &  {\scriptsize \textit{670, 1600, 300}} &  {\scriptsize \textit{700, 1800, 300}} \\
\hdashline
325 & 0.8 -- 8.0  & \textbf{1000} & \textbf{1100}\\
& &  {\scriptsize \textit{720, 1500, 200}} &  {\scriptsize \textit{780, 1600, 200}} \\
\hdashline
325 & 0.8 -- 15.0 & \textbf{1200} & \textbf{1200} \\
& &  {\scriptsize \textit{720, 1900, 300}} &  {\scriptsize \textit{780, 1600, 300}} \\
\hline
421.7 & 0.5 -- 4.5 & \textbf{800} & \textbf{900} \\
& &  {\scriptsize \textit{700, 1000, 100}} &  {\scriptsize \textit{700, 1100, 100}} \\
\hdashline
427.7 & 0.5 -- 4.5 & \textbf{1000} & \textbf{900} \\
& &  {\scriptsize \textit{800, 1200, 100}} &  {\scriptsize \textit{800, 1100, 100}} \\
\hdashline
448 & 0.5 -- 4.5 & \textbf{1500} & \textbf{1400} \\
& &  {\scriptsize \textit{1200, 1700, 100}} &  {\scriptsize \textit{1200, 1700, 100}} \\
\hline
630 & 0.8 -- 8.0  & \textbf{1600} & \textbf{1600}\\
& &  {\scriptsize \textit{1300, 2100, 100}} &  {\scriptsize \textit{1300, 2600, 200}} \\
\hdashline
630 & 0.8 -- 15.0 & \textbf{1800} & \textbf{1700} \\
& &  {\scriptsize \textit{1300, 2900, 300}} &  {\scriptsize \textit{1300, 2600, 200}} \\
\hdashline
650 & 0.8 -- 8.0  & \textbf{1800}& \textbf{1900} \\
& &  {\scriptsize \textit{1500, 2600, 100}} &  {\scriptsize \textit{1700, 2600, 80}} \\
\hdashline
650 & 0.8 -- 15.0 & \textbf{2300} & \textbf{2200}\\
& &  {\scriptsize \textit{1500, 4100, 700}} &  {\scriptsize \textit{1700, 2800, 300}} \\
\hline
\end{tabular}
\end{table}

\section{Discussion}
These results, presented on submillimeter-wave heterodyne Schottky diode receivers with integrated orthogonal-mode RF probe transitions, demonstrate leading performance and open new possibilities for the compact integration of high-frequency receivers. A summary comparison with polarimetric, compact receivers is provided in Table~\ref{TrecresultssumaryF}.  
The 325~GHz dual-polarization receiver exhibits a low DSB noise temperature across a wide 15~GHz IF bandwidth, consistent with previous results at 340~GHz \cite{Sobis2012}. Recent developments in InP HEMT low-noise amplifiers \cite{Kangaslahti2016} show superior noise performance; however, such devices may be more susceptible to temperature and temporal stability variations due to the intrinsic characteristics of the amplifier. The 424~GHz and 650~GHz receivers achieve state-of-the-art noise temperatures reported in the literature \cite{Treuttel2016} around 424~GHz and above 620~GHz, respectively, with performance maintained up to 8~GHz IF.  The integrated dual-polarization probes exhibit exceptionally low insertion loss, outperforming silicon micromachined devices and conventional CNC-machined OMTs, thereby minimizing the overall receiver noise temperature. The measured cross-polarization is above $20$~dB, consistent with state-of-the-art OMT performance at these frequencies and comparable to wire-grid polarizers (with typically below 1 dB insertion loss and 24 dB cross-polarization isolation between 602 and 740~GHz \cite{Realini2024} or less than 1~\% transmission loss and 34 dB cross polarisation isolation at 636 GHz\cite{Jacob2019}). Consequently, the presented concept simplifies the architecture of submillimeter dual-polarization receivers in terms of size and mechanical complexity without compromising polarimetric fidelity, such as insertion loss and cross-polarization. Further enhancements could be achieved by incorporating digital signal processing methods and utilizing three or more RF probes, as suggested by Morgan \textit{et al.} \cite{Morgan2010}. 

\begin{table}[h]
\centering
\caption{Comparison of compact submillimeter wave polarimetric receivers at room temperature}
\label{TrecresultssumaryF}
\begin{tabular}{|c|c|c|c|}
\hline
Freq. Band & IF & Receiver Noise Temperature &  Ref. \\
(GHz) & (GHz) & (K) &  \\
\hline
\textbf{315 - 325} & \textbf{0.8 - 15} & \textbf{720 - 1900 (1100 av. - DSB)}  & \textbf{This work} \\
\hline
\textbf{422 - 448} & \textbf{0.5 - 4.5} & \textbf{700 - 1700 (1100 av. - DSB)} & \textbf{This work} \\
\hline
660 - 680 & NA  & 3700 (Direct Detection) &  \cite{Cooke2020} \\
\hline
675 - 693 & 1 - 18 & 6000 (SSB) &  \cite{Bryerton2018}\\
\hline
\textbf{630 - 650} & \textbf{0.8 - 8} & \textbf{1300 - 2600 (1700 av. - DSB)}  & \textbf{This work} \\
\hline
\end{tabular}
\end{table}

\section{Conclusions}
State-of-the-art, compact, low-loss dual-polarization submillimeter receivers have been demonstrated, each equipped with two heterodyne subharmonic mixers and IF LNAs. The concept has been validated across three high-relevance Earth-observation passive remote sensing frequency bands: 325~GHz, 424~GHz, and 650~GHz. The integrated dual-polarization probe transitions show low insertion loss (below 0.5 dB) and high cross-polarization isolation (better than 20 dB) across all three bands. Double-sideband (DSB) receiver noise temperatures reached 830~K, 830~K, and 1620~K at 315~GHz, 421~GHz, and 630~GHz, respectively, with minimum measured DSB noise temperatures of 800~K, 800~K, and 1600~K. Each receiver offers a broad RF bandwidth and an overall gain of about 30 dB. These receivers, with state-of-the-art radiometric performance, can also be used in a single-polarization instrument, offering redundancy options. 

These advanced results in noise performance and polarization discrimination are achieved through precise integration. In summary, these developments pave the way for future atmospheric and weather satellite missions that require compact, sensitive polarimetric receivers. 

\section*{Acknowledgment}
The authors would like to thank Mats Myremark, Ahmed Salek Brzouami, and Jonathan Westin for machining the mechanical parts of the receiver and measurement setup, as well as the other people involved in the project: Dr. Arvid Hammar, Johanna Hanning, Simon Olvhammar, Dr. Slavko Dejanovic, Kalle Kempe, Rasmus Augustsson, Steve Sahlberg, and Martin Anderberg. The authors would also like to thank Elena Saenz, Martin van der Vost, and David Cuadrado-Calle at the European Space Agency for their input throughout the project's milestones. Finally, the authors would like to recognize Dr. Divya Jayasankar for her help in assembling the mixer substrate at 650~GHz. The integrated devices were fabricated and measured in the Nanofabrication Laboratory and the Kollberg Laboratory at Chalmers University of Technology in Gothenburg, Sweden.

\bibliographystyle{IEEEtran}
\bibliography{IEEEfull,bibl}

\begin{IEEEbiography}[{\includegraphics[width=1in,height=1.25in,clip,keepaspectratio]{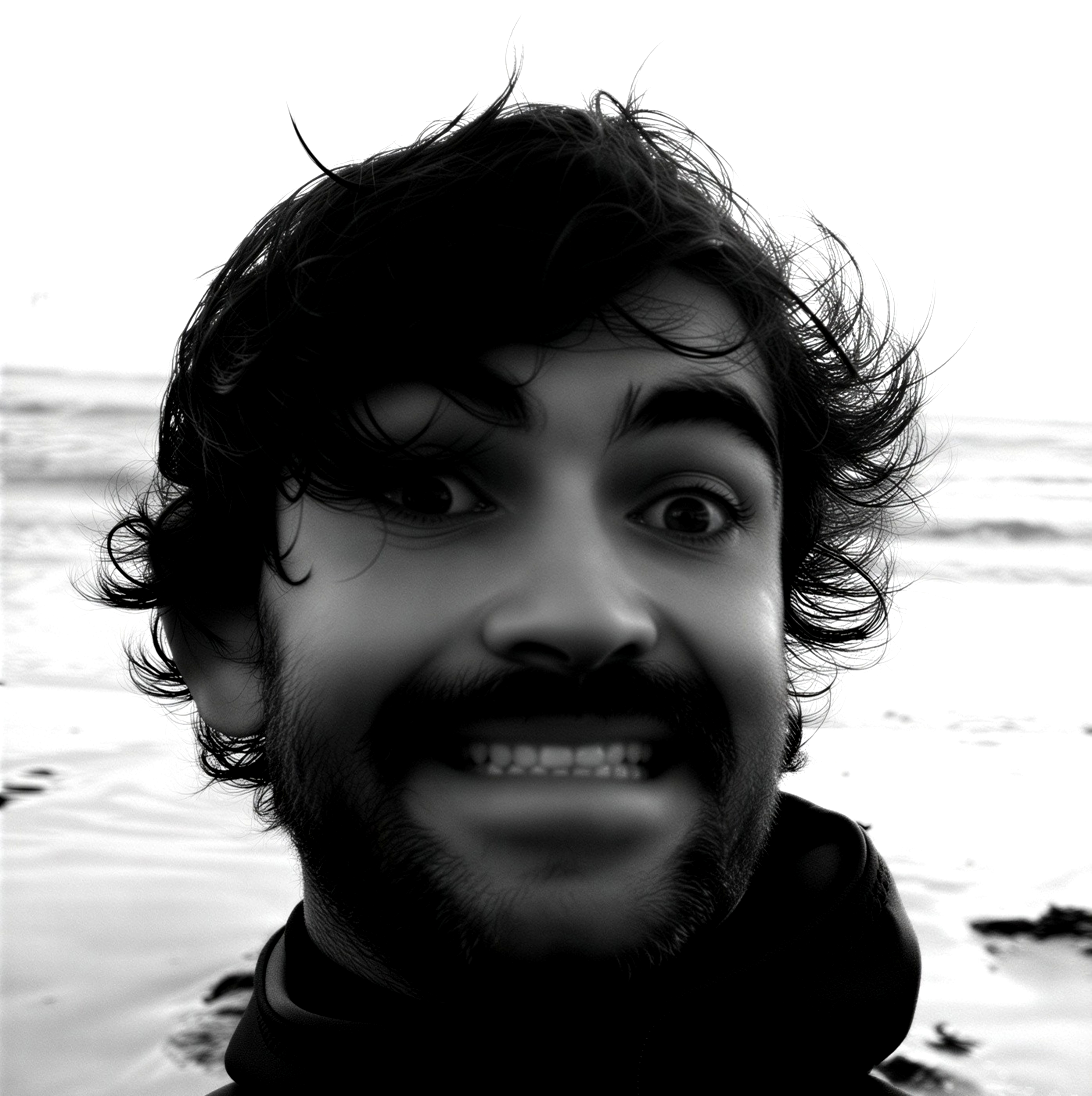}}]%
    {Olivier Auriacombe} was born in Vannes, France, in 1989. He received his M.Sc. degree in engineering and physics from Grenoble-INP, PHELMA, Grenoble, France, in 2013, and Ph.D degree in astronomy from the Open University, Milton Keynes, UK, in collaboration with the Rutherford Appleton Laboratory, Didcot, UK, in 2019. 
    
    From 2017 to 2019, he was involved with the development of radiometers and receivers (LOCUS, MetOp-SG) for the Millimetre-Wave Technology Group at the Rutherford Appleton Laboratory, UK. Between 2019 and 2023, he was postdoctoral research assistant at the Department of Microtechnology and Nanoscience (MC2) at Chalmers University of Technology, Gothenburg, Sweden. In 2021, he became project manager and RF engineer at the remote sensing satellite company, AAC Omnisys, part of AAC Clyde Space Group (working on receiver development, SKAO mid-band 1 receiver and the Arctic Weather Satellite). In September 2024, he joined MC2, Chalmers University of Technology as visiting researcher on high-frequency Schottky diode-based receivers. His current research involves Schottky-based THz systems, Radiometers, Radars, RF measurement techniques, THz spectroscopy, and Astrochemistry.
\end{IEEEbiography}

\begin{IEEEbiography}[{\includegraphics[width=1in,height=1.25in,clip,keepaspectratio]{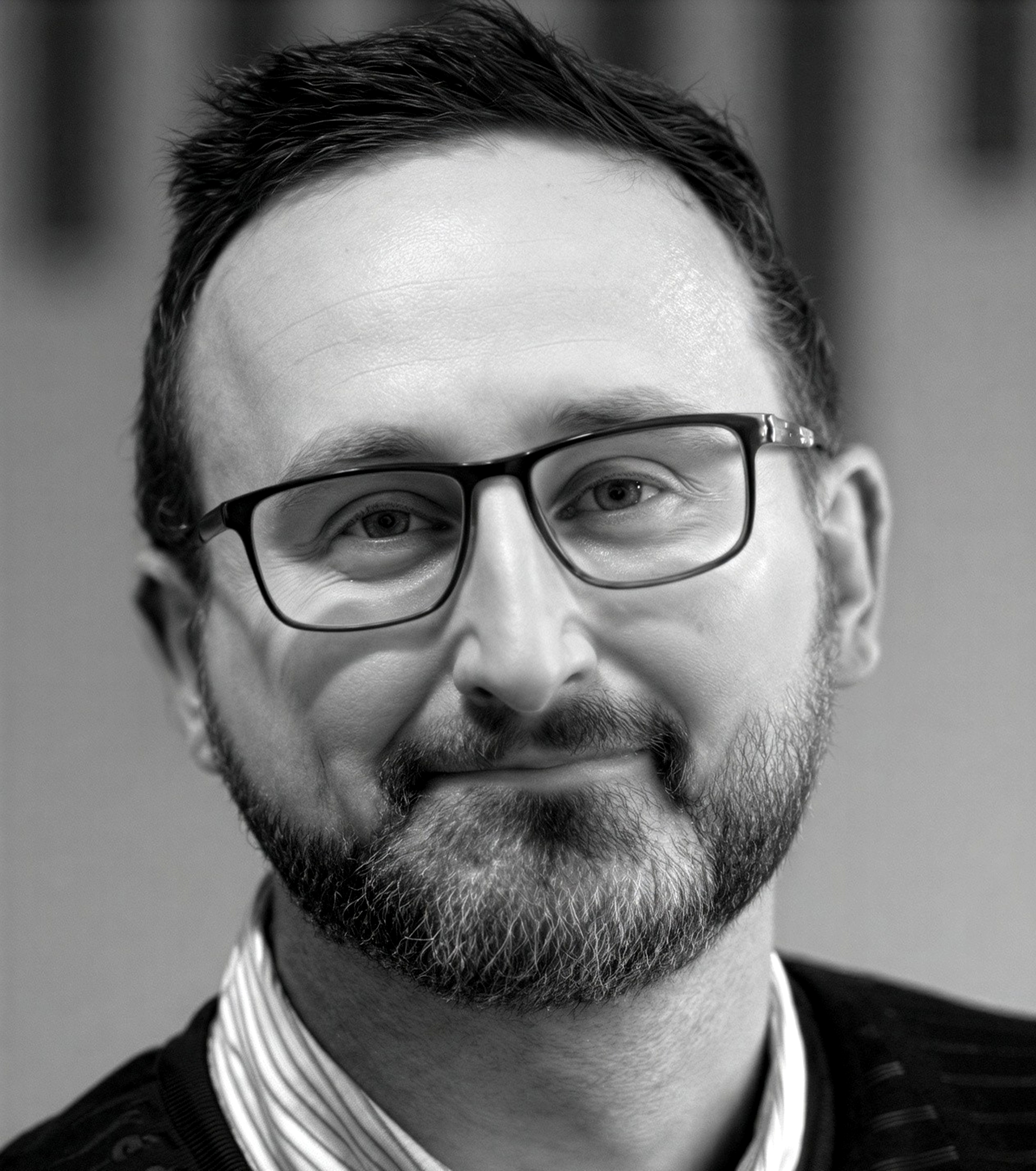}}]%
    {Peter J. Sobis} was born in Gothenburg, Sweden, in 1978. He received the M.Sc. degree in electrical engineering in 2003 and the Licentiate degree and Ph.D. in terahertz electronics from the Chalmers University of Technology, Gothenburg, Sweden, in 2010 and 2016, respectively. 
    
    From 2003 to 2004, he was with Anaren Microwave Inc., Syracuse, NY, USA, working on passive microwave components and beamforming networks. From 2004 to 2021, he was with Omnisys Instruments AB in Sweden, responsible for developing radiometer components and subsystems for various ESA and Swedish National Space Board Projects. In 2018, he became an Adjunct Professor in the Department of Microtechnology and Nanoscience (MC2) at Chalmers University of Technology. In 2021, he joined Low Noise Factory AB, Sweden, as a Senior RF Engineer. His current research and engineering work involves the development of micro- and mm-wave InP HEMT low-noise amplifiers and terahertz Schottky barrier diode components for quantum computer engineering, Earth observation instrumentation, and radio astronomy applications.
\end{IEEEbiography}

\begin{IEEEbiography}[{\includegraphics[width=1in,height=1.25in,clip,keepaspectratio]{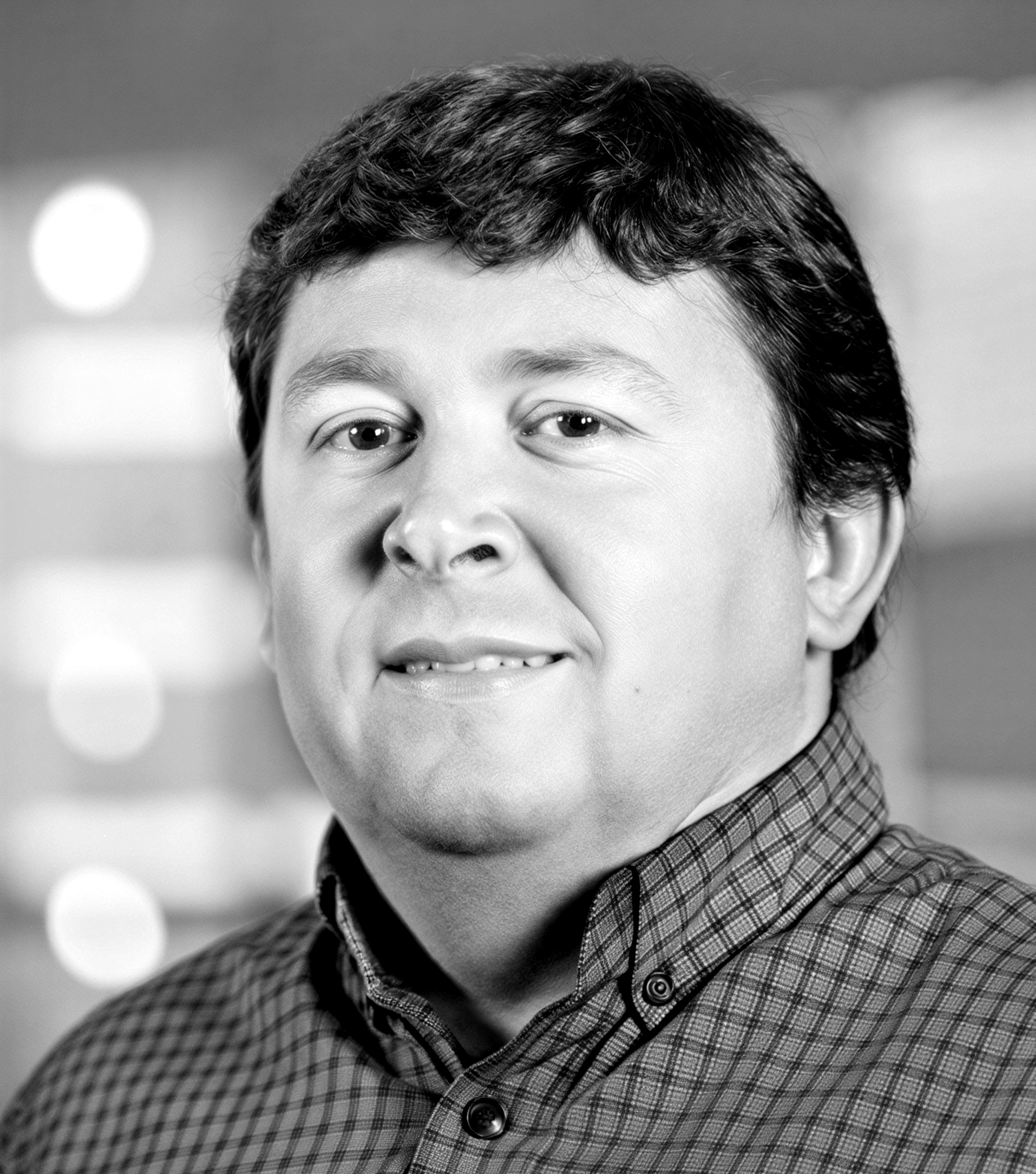}}]%
    {Vladimir Drakinskiy} was born in Kurganinsk, Russia, in 1977. He received a Diploma in Physics and Informatics (with honors) from the Armavir State Pedagogical Institute, Armavir, Russia, in 2000.
    
    From 2000 to 2003, he worked as a Junior Research Assistant at Moscow State Pedagogical University. Since 2003, he has been with the Department of Microtechnology and Nanoscience at Chalmers University of Technology, Gothenburg, Sweden, where he was responsible for fabricating Hot-Electron Bolometer terahertz mixer chips for the Herschel Space Observatory (2003–2005). Since 2008, he has been responsible for the terahertz Schottky barrier diode process line and was promoted to Senior Research Engineer in 2022. His research interests include micro- and nanofabrication techniques, submillimeter- and terahertz-detectors, and superconducting thin films.
\end{IEEEbiography}

\begin{IEEEbiography}[{\includegraphics[width=1in,height=1.25in,clip,keepaspectratio]{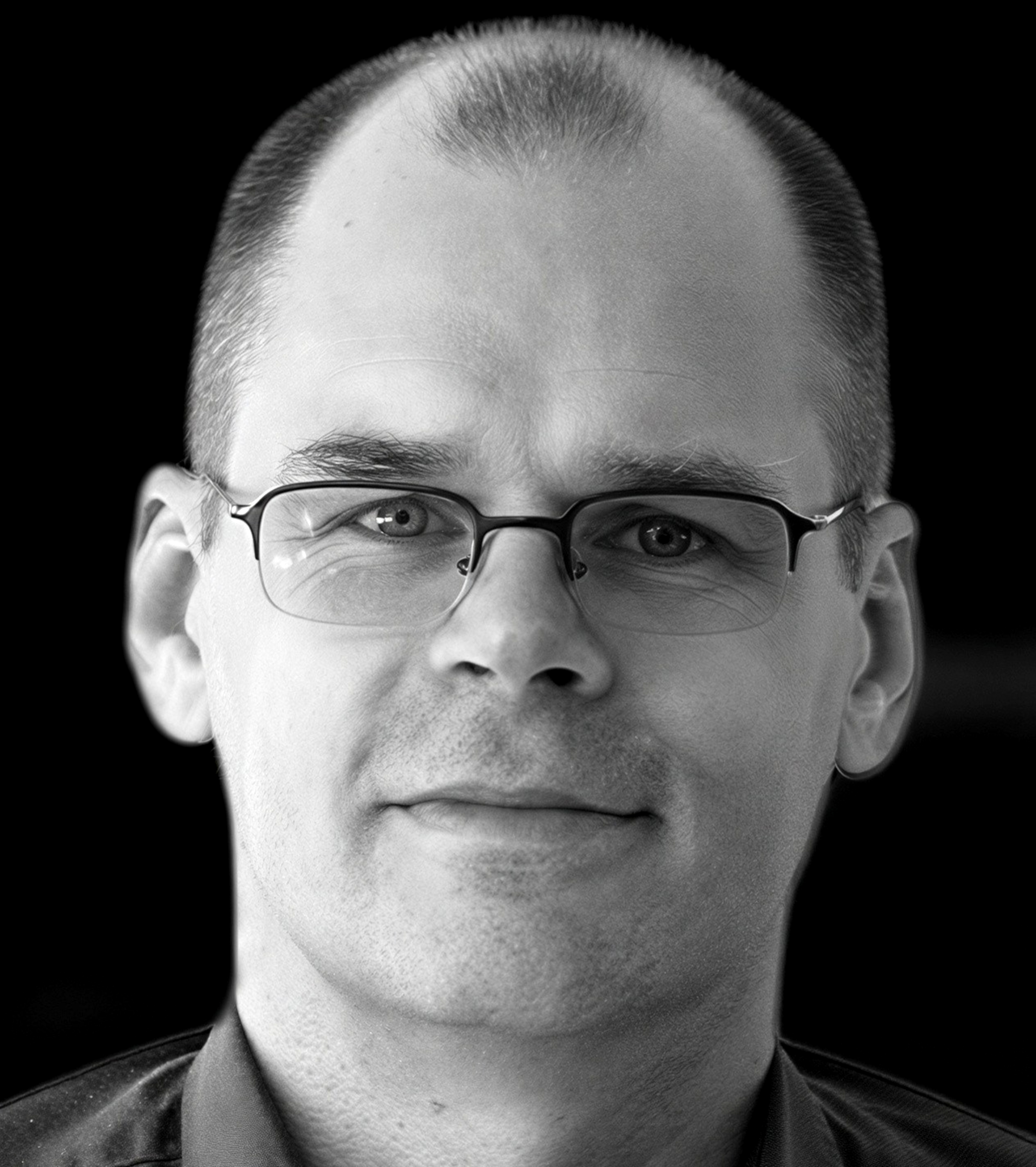}}]%
    {Anders Emrich} was born in Uddevalla, Sweden in 1962. He received his M.Sc. and Ph.D. degrees in electrical engineering from the Chalmers University of Technology, Göteborg, Sweden, in 1985 and 1992, respectively.
    
    In 1992, together with S. Andersson, he founded Omnisys Instruments AB, Västra Frölunda, Sweden, and was responsible for several subsystems for the ODIN radiometer payload and many development contracts toward ESA/ESTEC. He has been engaged in research collaborations with the Chalmers University of Technology and other research institutes and universities for more than 30 years. He has been responsible for the design, development, and production of the 183-GHz Water Vapor Radiometer (58 units) for ALMA, the GAS radio interferometer demonstrator (ESA contract), and the STEAMR instrument. He is currently leading the design and development of the instrument and several subsystems.
\end{IEEEbiography}

\begin{IEEEbiography}[{\includegraphics[width=1in,height=1.25in,clip,keepaspectratio]{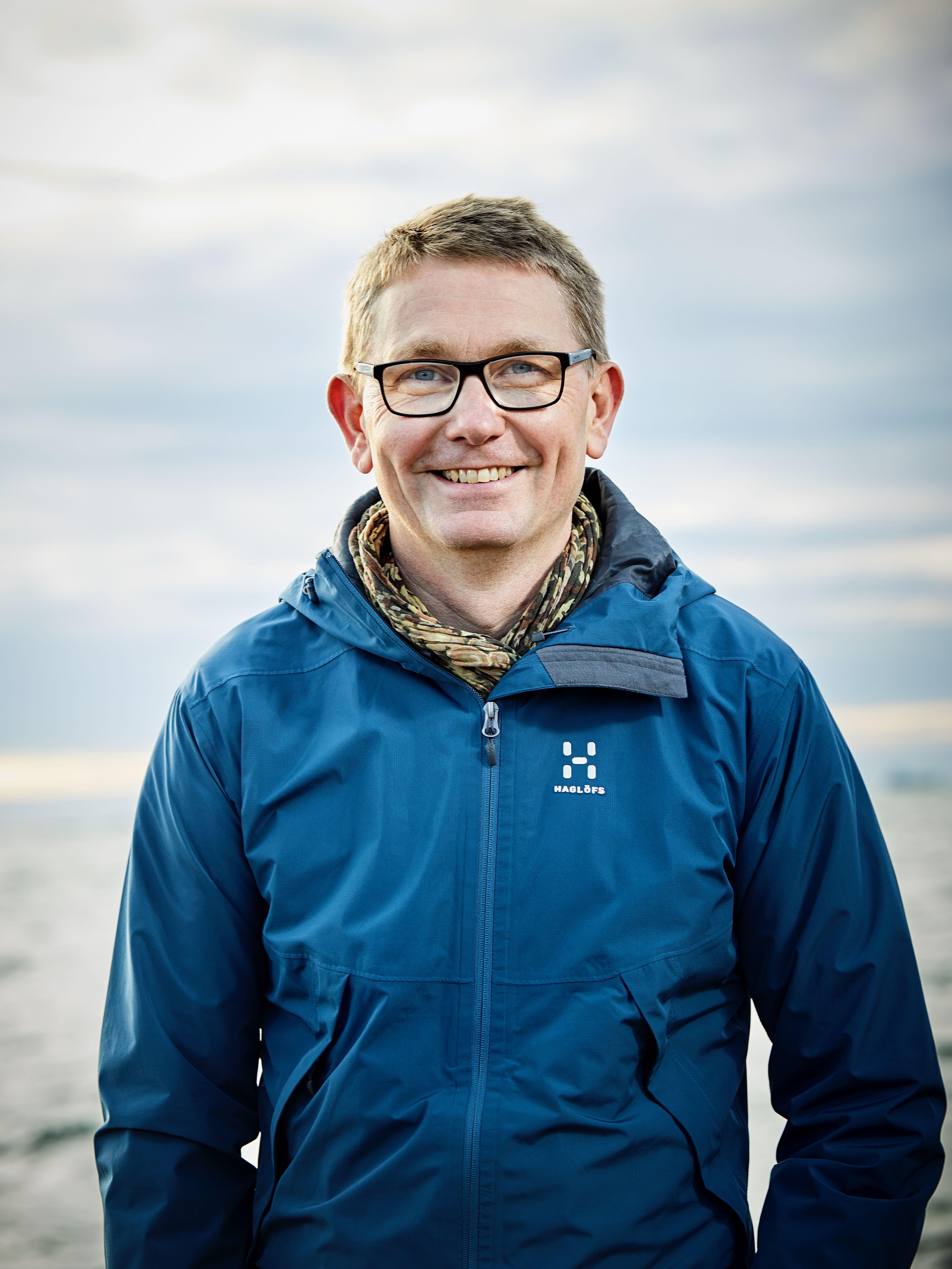}}]%
    {Jan Stake} (Fellow, IEEE) was born in Uddevalla, Sweden, in 1971. He received an M.Sc. in electrical engineering and a Ph.D. in microwave electronics from Chalmers University of Technology in Gothenburg, Sweden, in 1994 and 1999, respectively.
    
    In 1997, he was a Research Assistant at the University of Virginia, Charlottesville, VA, USA. From 1999 to 2001, he was a Research Fellow with the Millimetre Wave Group at the Rutherford Appleton Laboratory, Didcot, UK. He then joined Saab Combitech Systems AB, Gothenburg, Sweden, as a Senior RF/microwave Engineer until 2003. From 2000 to 2006, he held various academic positions at Chalmers University of Technology, and from 2003 to 2006, he also served as the Head of the Nanofabrication Laboratory in the Department of Microtechnology and Nanoscience (MC2). In 2006, he was appointed Professor and the Head of the Terahertz and Millimetre Wave Laboratory at Chalmers University of Technology. He was a Visiting Professor with the Submillimeter Wave Advanced Technology (SWAT) Group at Caltech/JPL, Pasadena, CA, USA, in 2007 and at TU Delft, the Netherlands, in 2020. He received an appointment as a Visiting Research Fellow at the National Physical Laboratory, UK, in 2024. Additionally, he is the co-founder of Wasa Millimeter Wave AB, Gothenburg, Sweden. His research interests include high-frequency semiconductor devices, terahertz electronics, submillimeter wave measurement techniques, and terahertz systems.
    
    Prof. Stake served as Editor-in-Chief of the IEEE Transactions on Terahertz Science and Technology from 2016 to 2018 and as Topical Editor from 2012 to 2015. From 2019 to 2021, he was chair of the IEEE THz Science and Technology Best Paper Award committee. He served on the International Society of Infrared, Millimeter, and Terahertz Waves (IRMMW-THz) organization committee from 2017 to 2024.
\end{IEEEbiography}
\end{document}